\newif\ifcheckpagelimits
\checkpagelimitstrue
\checkpagelimitsfalse

\ifcheckpagelimits
 \documentclass[nofootinbib,pra,aps,twocolumn,showpacs,showkeys,%
 amsmath,amssymb,superscriptaddress,final,reprint,floatfix,longbibliography]{revtex4-1}
 \newcommand{\todo}[1]{}
\else
 \documentclass[pra,aps,twocolumn,showpacs,showkeys,%
 amsmath,amssymb,superscriptaddress,final,reprint,floatfix,longbibliography]{revtex4-1}
 \newcommand{\todo}[1]{{\pdfmargincomment[icon=Note,color=pink]{#1}}}
\fi

\usepackage{lineno}
  \usepackage{mathptmx}
  \usepackage{amssymb}
\usepackage{amsmath}
\usepackage{amsfonts}
\usepackage{subfigure}
\usepackage{dcolumn}
\usepackage{amsmath,amssymb}
\usepackage{bm}
\usepackage{color}
\usepackage{overpic}
\usepackage{latexsym}
\usepackage{epstopdf}
\usepackage{color}
\usepackage[english]{babel}
\usepackage{latexsym}
\usepackage{stmaryrd}

\usepackage{braket}

\definecolor{mygrey}{gray}{0.35}
\definecolor{myblue}{rgb}{0.2,0.2,0.8}
\definecolor{myzard}{cmyk}{0,0,0.05,0}
\definecolor{mywhite}{rgb}{1,1,1}
\definecolor{myred}{rgb}{1,0.,0.3}

\usepackage[colorlinks=true,citecolor=myblue,linkcolor=myred]{hyperref}
\DeclareMathAlphabet{\mathpzc}{OT1}{pzc}{m}{it}

 \def\ee{\mathord{\rm e}}
 
 \def\ii{\mathord{\rm i}}

\def\half{\textstyle\frac{1}{2}}

\renewcommand{\ii}{{\rm i}}
\renewcommand{\ee}{{\rm e}}
\def\beq{\begin{equation}}
\def\eeq{\end{equation}}

\def\barray{\begin{eqnarray}}
\def\earray{\end{eqnarray}}

\begin{document}

\title{Quantum sensors for the generating functional of  interacting quantum field theories}

\author{A. Bermudez}
\email[Email to: ]{bermudez.carballo@gmail.com}
\affiliation{Department of Physics, College of Science, Swansea University, Singleton Park, Swansea SA2 8PP, United Kingdom}
\affiliation{Instituto de F\'isica Fundamental, IFF-CSIC, Madrid E-28006, Spain}

\author{G. Aarts}
\affiliation{Department of Physics, College of Science, Swansea University, Singleton Park, Swansea SA2 8PP, United Kingdom}

\author{M. M\"{u}ller}
\affiliation{Department of Physics, College of Science, Swansea University, Singleton Park, Swansea SA2 8PP, United Kingdom}

\begin{abstract}
 Difficult problems described in terms of interacting quantum fields
evolving in real time or out of equilibrium are abound in condensed-matter
and high-energy physics. Addressing such problems via controlled
experiments in atomic, molecular, and optical physics would be a
breakthrough in the field of quantum simulations. In this work, we present a quantum-sensing protocol to measure the generating functional of an interacting quantum field theory and, with it, all the relevant information about its in or out of equilibrium phenomena. Our protocol can be understood as a collective interferometric scheme based on a generalization of the notion of Schwinger sources in quantum field theories, which make it possible to  probe  the generating functional.  We show that our scheme can be realized in crystals of trapped ions acting as analog quantum simulators of self-interacting scalar quantum field theories.
  
\end{abstract}

\ifcheckpagelimits\else
\maketitle
\fi

 \ifcheckpagelimits
\else

\maketitle
\fi
\setcounter{tocdepth}{2}
\begingroup
\hypersetup{linkcolor=black}
\tableofcontents
\endgroup

\section{\bf Introduction}

Some of  the most complicated problems of  theoretical physics arise in the study of   quantum systems with a large,  sometimes  even infinite,  number of coupled  degrees of freedom. These complex problems arise in our effort to understand certain  observations in   condensed-matter~\cite{cm_book} or high-energy physics~\cite{qft_book}, which one tries to model with the unifying language of quantum field theories (QFTs). More recently, the field of atomic, molecular and optical (AMO) physics is  providing experimental setups~\cite{QS_cold_atoms,QS_trapped_ions} that aim at targeting similar problems. The approach is, however, rather different.   These AMO setups can be microscopically designed to behave with great accuracy according to a particular model of interest. Hence, it is envisioned that one will be capable of  answering open questions about a many-body model described through a QFT  by preparing, evolving, and measuring the experimental system;  what has been called a {\it quantum simulation}~\cite{feynman_qs,qs_goals}.

Either in the form of piecewise time evolution by concatenated unitaries~\cite{digital_QS}, i.e.\ digital quantum simulation (DQS), or continuous time evolution by always-on couplings~\cite{analog_QS}, i.e.\ analog quantum simulation (AQS), the main focus in this field has been typically placed   on the quantum simulation of condensed-matter problems~\cite{QS_cold_atoms,QS_trapped_ions,qs_book}. Nonetheless, some theoretical works  have also addressed  how quantum simulators could  mimic the relativistic QFTs that appear in high-energy physics, as occurs for the AQS of a  Klein-Gordon QFT with Bose-Einstein condensates~\cite{aqs_qft_bosons_bec,aqs_qft_bosons_bec_detection}. Note, however, that the most versatile AMO quantum simulators to date~\cite{QS_cold_atoms,QS_trapped_ions}  do not work directly in the continuum, but on a physical lattice that is either provided by additional laser dipole forces for neutral atoms~\cite{QS_cold_atoms}, or by the interplay of Coulomb repulsion and electromagnetic oscillating forces for singly-ionized atoms~\cite{QS_trapped_ions}. Therefore,  the relevant symmetries of the high-energy QFT, such as Lorentz invariance, must emerge as one takes the continuum/low-energy limit in the AMO quantum simulator. This occurs trivially for free fermionic QFTs~\cite{comment_free_fermions}, which underlies the schemes for the AQS of   Dirac QFTs with ultra-cold atoms in optical lattices~\cite{aqs_qft_fermions}. There are also proposals for interacting QFTs, such as  the DQS of self-interacting Klein-Gordon fields~\cite{dqs_qft_phi4}, the analog~\cite{aqs_qft_boson_fermion} and digital~\cite{dqs_qft_boson_fermion} quantum simulators of coupled Fermi-Bose fields, and an ultra-cold atom AQS of  Dirac  fields with self-interactions or coupled to scalar bosonic fields~\cite{aqs_qft_fermions_interactions}. 

In the interacting case, as discussed in~\cite{aqs_qft_fermions_interactions,dqs_qft_phi4}, renormalization techniques must be employed  to set the right bare parameters in  such a way that  a QFT with the required Lorentz symmetry and free of ultraviolet (UV) divergences is obtained in the continuum limit. This is the standard situation in  {\it lattice field theories}~\cite{book_lattice}, where the continuum limit  is obtained by letting the lattice spacing $a\to 0$,    removing thus the natural UV cut-off of the  lattice,    while maintaining a finite renormalized mass/gap $m$ describing the physical mass of the particles in the corresponding QFT. This requires setting the bare parameters close to a critical point of the lattice model, where  the dimensionless correlation length, measured   in lattice units,  diverges $\tilde{\xi}\to\infty$. In this case,  the  mass  $m\sim1 /\tilde{\xi} a$ can remain constant even for  vanishingly small lattice spacings. Therefore, the experience gained in the classical numerical simulation of interacting QFTs on the  lattice will be of the utmost importance for the progress  of quantum simulators of high-energy physics problems.

In a more direct connection to open problems in high-energy physics, e.g. the phase diagram of  quantum chromodynamics~\cite{qcd_phase_diagram}, we note that there has been a number of proposals for the DQS~\cite{dqs_lattice_gauge_theories,dqs_qft_gauge_U(1)_qlinks,dqs_lattice_gauge_theories,dqs_qft_SU(2)_qlinks} and AQS~\cite{aqs_qft_gauge_U(1)_qlinks,aqs_QED,aqs_qft_gauge_U(1)_qlinks,aqs_qft_fermions_U(1)_qlinks,aqs_qft_fermions_SU(2)_qlinks,aqs_qft_fermions_U(1)_qlinks,aqs_qft_fermions_SU(2)_qlinks} of gauge  theories. As announced above, previous knowledge from lattice gauge theories has been essential  to come up with schemes for the quantum simulation of Abelian~\cite{aqs_qft_gauge_U(1)_qlinks,dqs_qft_gauge_U(1)_qlinks} and non-Abelian~\cite{dqs_qft_SU(2)_qlinks} QFTs of the gauge sector,  as well as Abelian~\cite{aqs_QED,aqs_qft_fermions_U(1)_qlinks} and non-Abelian~\cite{aqs_qft_fermions_SU(2)_qlinks} QFTs of gauge fields coupled to Dirac fields. Starting from the simpler QFTs discussed above, this body of work  constitutes a well-defined long-term roadmap for the implementation of relevant models of high-energy physics in AMO platforms~\cite{zohar_review}. In this work, we address  the question of devising a general measurement strategy   to extract the properties of an interacting QFT, which could be adapted to  these different quantum simulators. One possibility would be to mimic the high-energy scattering experiments in particle accelerators by preparing wave packets and measuring the outcome after a collision, as proposed  in the context of DQS~\cite{dqs_qft_phi4}.  In this work, we explore a different possibility  that would allow the quantum simulator to extract the complete information about an interacting QFT. 
We introduce a  scheme that is capable of measuring the  {\it generating functional} of the QFT~\cite{qft_book}. In particular, this functional can be used  to extract the {\it Feynman propagator}, such that one can also make predictions about different scattering experiments. In addition, other relevant properties of the interacting QFT can also be directly extracted from such a functional. Moreover, our scheme is devised for analog quantum simulators, such that the resource requirements are  lower than those of a DQS using a fault-tolerant quantum computing hardware.

\begin{figure*}
\centering
\includegraphics[width=1.7\columnwidth]{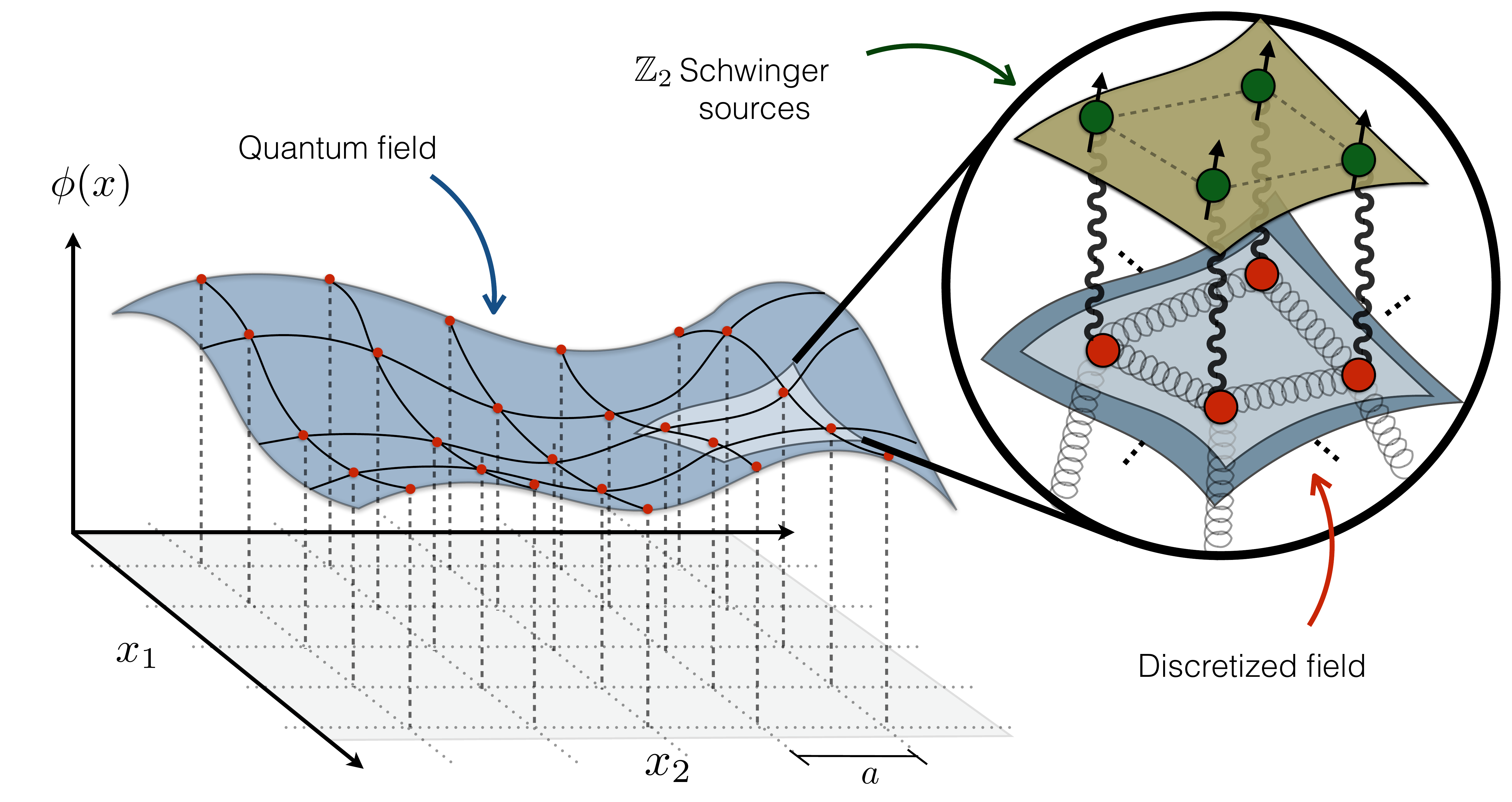}
\caption{ {\bf Schematic representation of the Schwinger sensors for  the generating functional:} We represent a quantum scalar field $\phi(x)$ in a $D=2+1$ space-time, which is discretized into a $d=2$ spatial lattice, while letting the time coordinate  continuous. (inset) Zoom of a small space region, where the field at each point  (red circles) is coupled to the fields at neighbouring points (small springs), and can be excited by its coupling (wavy lines) to  generalised quantum-mechanical Schwinger sources (green circles with  arrows). These $\mathbb{Z}_2$ Schwinger sources  will also function as  quantum sensors for the generating functional.}
\label{fig_field_sensing}
\end{figure*}

\section{\bf  Sensors for  quantum field theories (QFTs)}
\label{sec:sensors_qft}
In this section, we introduce a scheme to measure the generating functional of a QFT directly in the continuum. For the sake of concreteness, we present our results by focusing on a real scalar QFT, and comment on generalizations to other QFTs at the end of the section.

\subsection{Self-interacting  Klein-Gordon QFT, Schwinger sources, and the generating functional}

Let us consider a self-interacting real Klein-Gordon QFT, which is described by the bosonic scalar field operator $\phi(x)$, where $x=(t,\bf{x})$ is a point in the $D=(d+1)$-dimensional Minkowski space-time with coordinates $x^\mu$, $\mu\in\{0,\cdots,d\}$, and we set $\hbar=c=k_{\rm B}=1$. The Lagrangian 
density that governs the dynamics of the scalar field is
\beq
\label{eq:int_lag}
\mathcal{L}=\frac{1}{2}\partial_\mu\phi(x)\partial^\mu\phi(x)-\frac{m_0^2}{2}\phi(x)^2-\mathcal{V}(\phi),
\eeq
where $\partial_\mu=\partial/\partial x^\mu$,  $\partial^\mu=\eta^{\mu\nu}\partial_\nu$ with Minkowski's  metric $\eta={\rm diag}(1,-1,\cdots,-1)$, and we use Einstein's  summation criterion for repeated indexes. Here,  $m_0$ is the bare mass of the scalar boson, and $\mathcal{V}(\phi)$ describes its self-interaction through non-linearities e.g. $\lambda\phi^4$ or $\cos(\beta\phi)$. In these units, in order to make the action  $S=\int {\rm d}^Dx\mathcal{L}$ dimensionless, the scalar field must have classical mass dimensions $d_\phi=(D-2)/2$, while the couplings have $d_{m_0^2}=2$ and $d_{\lambda}=(4-D)$.

 Let us now introduce the so-called Schwinger sources~\cite{schwinger}, which are classical background fields that generate excitations of the quantum field. For the real scalar QFT~\cite{qft_book}, it suffices to introduce a classical scalar background field $J(x)$ and modify the Lagrangian according to  
\beq
\label{eq:schwinger_sources}
\mathcal{L}\to\mathcal{L}_J=\mathcal{L}+J(x)\phi(x),
\eeq
where the  sources have mass dimension $d_J=(D+2)/2$. The normalized generating functional is obtained from the vacuum-to-vacuum propagator, after removing  processes where particles are spontaneously created/annihilated  in the absence of the Schwinger sources. This can be expressed as 
\beq
\label{eq:generating_functional}
\mathsf{Z}[J(x)]=\bra{\Omega}\mathsf{T}\!\left\{\ee^{\ii\int {\rm d}^DxJ(x)\phi_{H}(x)}\right\}\ket{\Omega},
\eeq
where we have introduced the ground state of the interacting QFT $\ket{\Omega}$,  and used the time-ordering symbol $\mathsf{T}$. Additionally,  the field operators are expressed in the Heisenberg picture of the interacting  QFT in the absence of Schwinger sources. This is achieved by defining  $\phi_{H}(x)=\mathsf{T}\{\ee^{\ii\int {\rm d}^Dx \mathcal{H}}\}\phi(x)\mathsf{T}\{\ee^{-\ii\int {\rm d}^Dx \mathcal{H}}\}$, where  the integral in the
evolution operator includes integration over time, and 
\beq
\label{eq:Ham}
\mathcal{H}=\frac{1}{2}\pi(x)^2+\frac{1}{2}\boldsymbol{\nabla}\phi(x)^2+\frac{m_0^2}{2}\phi(x)^2+\mathcal{V}(\phi)
\eeq
 is the Hamiltonian density associated to the QFT under study~\eqref{eq:int_lag}. Here, $\pi(x)=\partial_t\phi(x)$ is the conjugate momentum fulfilling the equal-time canonical commutation relations with the scalar field  $[\phi(t,{\bf x}),\pi(t,{\bf y})]=\ii \delta^d({\bf x}-{\bf y})$. 
 
 The normalized generating functional, hereafter simply referred to as  the generating functional,  contains all the relevant information about the QFT. In particular, any $n$-point Feynman propagator $G^{(n)}=\bra{\Omega}\mathsf{T}\left\{\phi_{H}(x_1)\cdots \phi_{H}(x_n)\right\}\ket{\Omega}$ can be obtained from $\mathsf{Z}[J]$ by functional differentiation 
 \beq
 \label{eq:n-part_propagator}
 G^{(n)}(x_1,\cdots,x_n)=(-\ii)^n\left.\frac{\delta^n\mathsf{Z}[J(x)]}{\delta J(x_1)\cdots\delta J(x_n)}\right|_{J=0}.
 \eeq
 Note that we are using the normalized   generating functional~\eqref{eq:generating_functional}, such that the factor $\mathsf{Z}^{-1}(0)$ in the propagator~\eqref{eq:n-part_propagator} disappears as $\mathsf{Z}(0)=1$.
 Through the Gell-Mann-Low theorem~\cite{gellman_low}, one can express the generating functional, and thus any $n$-point propagator of the interacting QFT, in terms of Feynman diagrams. Accordingly, $\mathsf{Z}[J]$ becomes a fundamental tool in the  theoretical study of interacting QFTs. The question that we address in the following subsection is if such a functional can also become an observable in some experiment. Note that we are not referring to susceptibilities expressed in terms of retarded Green's functions, which are typically measured in  linear-response experiments. We are instead looking for a  scheme that allows one to measure the complete generating functional, out of which one could calculate any time-resolved Feynman propagator, or obtain predictions of any type of scattering experiment. The generating functional does indeed contain all the relevant information about a QFT.

\subsection{$\mathbb{Z}_2$ Schwinger sources}

The proposed scheme  promotes the classical Schwinger fields~\eqref{eq:schwinger_sources} to quantum-mechanical $\mathbb{Z}_2$ Schwinger sources. In particular, we will consider the $\mathfrak{su}(2)$ Lie algebra,  and define the operators $\sigma^\alpha$, where $\sigma^0=\mathbb{I}$, and  $\{\sigma^ \beta\}_{\beta=1,2,3}$ are the well-known Pauli matrices. The $\mathbb{Z}_2$  Schwinger field now reads
\beq
\label{eq:upgrade}
J(x)\to \sum_\alpha J^{\alpha}(x)\sigma^\alpha(x),
\eeq
where $J^{\alpha}(x)$ are classical background fields, and $\sigma^\alpha(x)$ can be interpreted as the operators of an ancillary two-level system (i.e.\ spin-1/2, qubit) that is attached to every space-time coordinate. We advance, however, that for AQS of QFTs on the lattice, we shall not need a continuum but a countable set of ancillary spins/qubits (see Fig.~\ref{fig_field_sensing}). 

The classical background field~\eqref{eq:schwinger_sources}, which was introduced by Schwinger as a mathematical artefact in order to calculate the  generating functional of the interacting QFT~\eqref{eq:generating_functional}, has now been promoted onto a quantum-mechanical source that may also have its own dynamics described by a generic Hamiltonian $\mathcal{H}_\sigma$. Hence,  Eq.~\eqref{eq:Ham} must be substituted by
\beq
\label{eq:non_ab_sources}
\mathcal{H}\to\mathcal{H}_J=\mathcal{H}+\mathcal{H}_\sigma-\sum_\alpha J^\alpha(x)\sigma^\alpha(x)\phi(x).
\eeq
The main idea is that these quantum sources will not only act as generators of excitations in  the quantum field, but also as quantum probes capable of measuring the   generating functional of the interacting QFT. We discuss below a particular measurement protocol to achieve this goal.

The use of quantum-mechanical two-level systems as sensors for measuring physical quantities with high precision, such as electric/magnetic fields or oscillator frequencies, is a well-developed technique in AMO physics~\cite{q_sensors_review}. In the two most standard cases, the two-level system can get excited (i.e.\ Rabi probe) or gain a relative phase (i.e.\ Ramsey probe) as a consequence of its coupling to the physical quantities  that  need to be measured. In many situations of experimental relevance, one uses a single quantum sensor, and maintains its quantum coherence for ever-increasing periods of time to improve the sensitivity of the measurement apparatus. In the context of QFTs,   ever since the pioneering work of  W. G. Unruh~\cite{unruh_detector},  Rabi-type  probes  based on a single particle with discrete energy levels have been routinely considered as  detectors of quantum fields~\cite{book_cureved_qft}. These type of detectors have also been considered in a quantum-simulation context~\cite{aqs_qft_bosons_bec_detection,aqs_qft_effects_ions}.  In contrast, Ramsey-type probes have been mainly considered for the quantum simulation of condensed-matter problems (see e.g.~\cite{aqd_bec}). On the other hand, in the context of high-energy
physics, Ramsey probes remain largely unexplored (an exception is Ref.~\cite{zohar_ramsey}, which discusses interferometric measurements of string tension and
the Wilson-loop operator in gauge theories). In this work, we partially fill  this gap by showing that the interacting relativistic  QFT~\eqref{eq:non_ab_sources}  with $\mathfrak{su}(2)$ Schwinger sources of the Ramsey type, i.e.\ setting $J^\alpha(x)=J(x)(\delta_{\alpha,0}-\delta_{\alpha,3})/2$, can  function as a quantum sensor for the QFT generating functional. This particular choice of the generalised Schwinger sources guarantees a differential coupling of the field to the internal states of the sensors, such that their relative phase will depend on the time-evolution of the quantum scalar field. In the following we will show that a  collective interferometric scheme can exploit this differential coupling to probe the full generating functional.

\subsection{Quantum sensors for  the generating functional}

In addition to exploiting the quantization of energy levels and the quantum coherence, quantum sensors based on ensembles of two-level systems can make use of entanglement to increase their sensitivity~\cite{entanglement_metrology}, or to gain information about  equal-time density-density correlations from their short-time dynamics~\cite{density_correlations}, which can be of interest in the quantum simulation of condensed-matter problems. In our case, entanglement is not  used to increase the sensitivity, but is also an ingredient of paramount importance to map the whole information of the relativistic QFT, which is encoded in the generating functional $\mathsf{Z}[J]$, into the ensemble of $\mathbb{Z}_2$ Schwinger sources/sensors. Let us describe in detail the protocol.

We consider an initial state in the remote past as the tensor product of the  interacting QFT ground state  $\ket{\Omega}$ and the quantum sensor  state with all spins pointing down $\ket{0_\sigma}$. This assumes that the state of the quantum field is adiabatically prepared in the remote past by starting from the non-interacting ground state, and switching on the self-interactions $V(\phi)$ sufficiently slow, i.e.\ adiabatically, while the Schwinger-source couplings remain switched off.
One then applies a fully entangling operation to the sensors generating the so-called Greenberger-H{o}rne-Zeilinger (GHZ) states~\cite{ghz}, which are multi-partite generalizations of the Einstein-Podolski-Rosen (EPR) states~\cite{epr}. This leads to 
\beq
\label{eq:ghz_all}
\ket{\Psi(t_0)}=\ket{\Omega}\otimes\frac{1}{\sqrt{2}}\left(\prod_{{\bf x}}\sigma^0(t_0,{\bf x})+\prod_{{\bf x}}\sigma^1(t_0,{\bf x})\right)\ket{0_\sigma}.
\eeq
  At this instant of time $t_0$, the Schwinger-source couplings are switched on. The time-evolution operator of the full sourced QFT~\eqref{eq:non_ab_sources} can be expressed in the interaction picture with respect to $\mathcal{H}_0(x)=\mathcal{H}+\mathcal{H}_\sigma$, being $\mathcal{H}$ the Hamiltonian of the sourceless interacting QFT~\eqref{eq:Ham}. In the distant future, the time-evolution operator becomes
\beq
\label{eq:time_evolution}
U_J=\mathsf{T}\left\{\ee^{-\ii \int {\rm d}^Dx\mathcal{H}_0(x)}\right\}\mathsf{T}\left\{\ee^{+\ii \int {\rm d}^DxJ(x)\phi_{\rm H}(x)P_{\rm H}(x)}\right\},
\eeq 
where $P_{\rm H}(x)=\mathsf{T}\{\ee^{\ii\int{\rm d}^Dx\mathsf{H}_\sigma(x)}\}P(t_0,{\bf x})\mathsf{T}\{\ee^{-\ii\int{\rm d}^Dx\mathcal{H}_\sigma(x)}\}$, and $P(t_0,{\bf x})=\half(\sigma^0({\bf x})-\sigma^3({\bf x}))$  is an orthogonal projector onto the ground state of the $\mathbb{Z}_2$ sensor localized at coordinate $(t_0,\bf{x})$. Finally, the observable  information about the generating functional will be encoded in the  expectation value of a spin parity operator
\beq 
\label{eq:full_parity}
\mathsf{P}[J(x)]= \bra{\Psi(t_0)}U_J^\dagger\prod_{{\bf x}}\sigma^1(t_0,{\bf x})U_J\ket{\Psi(t_0)}.
\eeq
We will consider a simple Hamiltonian for the quantum sensors 
\beq
\mathcal{H}_\sigma=\delta\epsilon(\sigma^0(t_0,{\bf x})-P(t_0,{\bf x})),
\eeq
 where $\delta\epsilon$ is the energy-density associated to the transition frequency $\omega_0$ between the two  levels of the sensor, which has a natural realization in quantum-simulation AMO platforms. For this particular choice, one finds that the expectation value of the parity operator evolves according to $\mathsf{P}[J]=\bra{\Omega}\mathsf{T}\left\{\ee^{\ii\int{\rm d}^Dx (\delta\epsilon+ J(x)\phi_{\rm H}(x))}\right\}\ket{\Omega}+{\rm c.c.}$, and thus encodes the generating functional~\eqref{eq:generating_functional}, namely 
\beq
\label{eq:parity_generating}
\mathsf{P}[J(x)]=\half\ee^{\ii\int{\rm d}^Dx\delta\epsilon }\mathsf{Z}[J(x)]+{\rm c.c.}
\eeq
Let us recapitulate the results obtained so far. By introducing a quantum sensor $\sigma^{\alpha}(x)$ at each space-time coordinate, thus upgrading the standard Schwinger sources  to $\mathbb{Z}_2$  fields~\eqref{eq:upgrade}, we have constructed a parity Ramsey interferometer  capable of encoding the  generating functional of an interacting QFT in its time-evolution~\eqref{eq:parity_generating} for a particular set of background sources fulfilling $J^\alpha(x)=J(x)(\delta_{\alpha,0}-\delta_{\alpha,3})/2$. 

\subsection{Simplified  sensors for  Feynman propagators}

In this section, we show that the protocol can be simplified considerably if one is only interested in  $n$-point Feynman propagators~\eqref{eq:n-part_propagator}. Such propagators contain a lot of information relevant to the typical scattering experiments and other types of real-time non-equilibrium phenomena.
For even $n$,  these Feynman  propagators can be inferred from  the Ramsey parity signal by functional differentiation
\beq
\label{eq:functional_derivatives_parity}
\left.\frac{\delta^n\mathsf{P}[J(x)]}{\delta J(x_1)\cdots\delta J(x_n)}\right|_{J=0}=\half\ee^{\ii\int{\rm d}^Dx\delta\epsilon }
G^{(n)}(x_1,\cdots,x_n)+{\rm c.c.},
\eeq
where we have assumed that an approximate remote-past to distant-future time-evolution is obtained by setting $t-t_0>{\rm max}\{|x_i^0-x_j^0|,\hspace{1ex}\forall\hspace{0.5ex} i,j=1,\cdots ,n\}$. 

To estimate such functional derivatives, the required Schwinger field $J(x)$ that must be experimentally applied would be a comb of $n$ point-like  sources
\beq
\label{eq:source_n_point}
J(x)=\sum_{i=1}^n \mathsf{J}_i\delta^D(x-x_i),
\eeq 
where $\mathsf{J}_i$ are the strengths of  infinitesimal field-sensor couplings at the particular space-time coordinates $x_i$. Since the Schwinger field~\eqref{eq:source_n_point} is only applied to a subset of the quantum sensors located at $X_{\rm s}=\{{\bf x}_1\cdots{\bf x}_n\}$, which already   requires addressability, we may also consider that the initial entangling operation may only involve that particular  subset. In that case,  one can simplify the required initial state~\eqref{eq:ghz_all} to 
\beq
\label{eq:partial_ghz}
\ket{\Psi(t_0)}=\ket{\Omega}\otimes\frac{1}{\sqrt{2}}\left(\prod_{{\bf x}\in X_{\rm s}}\sigma^0(t_0,{\bf x})+\prod_{{\bf x}\in X_{\rm s}}\sigma^1(t_0,{\bf x})\right)\ket{0_\sigma}.
\eeq
 The fact that a GHZ state of all the spins is no longer required is a very important simplification, and also makes the protocol more robust as one considers the degrading effect of external sources of noise. Additionally, we do not require to measure the full spin parity~\eqref{eq:full_parity}, but only 
\beq
\label{eq:partial_parity}
 \mathsf{P}[J(x)]= \bra{\Psi(t_0)}U_J^\dagger\prod_{{\bf x}\in X_{\rm s}}\sigma^1(t_0,{	\bf x})U_J\ket{\Psi(t_0)}. 
\eeq

Note that  mobile sensors might be available depending on the particular implementation. In that case, we do not require a quantum sensor at every space-time coordinate, but only   $n$ sensors located at the corresponding points $\mathcal{H}_\sigma=\sum_{j=1}^n\delta\epsilon(\sigma^0(t_0,{\bf x})-P(t_0,{\bf x}))\delta^d({\bf x}-{\bf x}_j)$. 

To infer the value of the functional derivative of the parity signal~\eqref{eq:partial_parity}, one  needs to apply different sets of instantaneous sources~\eqref{eq:source_n_point}, which we label by ${\mathbf{J}}^{(m)}=\left(\mathsf{J}_1^{(m)},\cdots,\mathsf{J}_n^{(m)}\right)$ with $m=1,\cdots,M$. For each of these sets of sources, one would then measure the corresponding parity oscillations   $\mathsf{P}[{\mathbf{J}}^{(m)}]$. Finally, by adding and subtracting  these parities according to a given prescription obtained by the discretization of the functional derivatives, one can infer an estimate of the $n$-point Feynman propagators via Eq.~\eqref{eq:functional_derivatives_parity}.

 To be more concrete, let us consider the important case of the single-particle 2-point Feynman propagator $\Delta(x_1-x_2)=G^{(2)}(x_1,x_2)$. In this case, our protocol requires creating a simple EPR pair between two distant quantum sensors at ${{\bf x}_1,{\bf x}_2}$~\eqref{eq:partial_ghz}. Additionally, 
 we have to consider  $M=4$  different measurements of the Ramsey parity signal for a time longer than $|x_1^0-x_2^0|$ with the following sets of Schwinger sources: ${\mathbf{J}}^{(1)}=(0,0)$, ${\mathbf{J}}^{(2)}=({\mathsf{J}}_{1},0)$, $\boldsymbol{{\mathbf{J}}^{(3)}}=({\mathsf{J}}_1, {\mathsf{J}}_2)$, and ${\mathbf{J}}^{(4)}=(0,{\mathsf{J}}_{2})$. Using these sets of infinitesimal Schwinger sources, one can reconstruct the discretization of the two functional derivatives required to calculate the single-particle Feynman propagator in Eq.~\eqref{eq:n-part_propagator} for $n=2$. Therefore,  the Feynman propagator can  then be inferred from
\beq
\label{two_point_propagator}
\sum_{m=1}^4\frac{(-1)^{m}\mathsf{P}[{\mathbf{J}}^{(m)}]}{{\mathsf{J}}_{1}{\mathsf{J}}_{2}}\approx-\ee^{\ii 2\omega_0 (t-t_0)}
\Delta(x_1-x_2)+{\rm c.c.},
\eeq
where we have assumed that ${\mathsf{J}}_1,{\mathsf{J}}_2\ll m_0^2$, and where $m_0$ is the bare mass of the QFT~\eqref{eq:int_lag}. According to this expression, we can infer the real (imaginary) part of the propagator by measuring at $\tau=2\pi r/\omega_0$ ($\tau=(2r+1)\pi /2\omega_0$), where $r\in\mathbb{Z}$. 

Let us now advance on the results of the following section, where we discuss an implementation of this sensing scheme using AMO quantum simulators of QFTs. In this case,  the quantum sensors  can also have spurious couplings to other quantum/classical fields, e.g. environmental electromagnetic fields, which cannot be switched on/off, but instead act continuously during the probing protocol. Accordingly, the parity oscillations will also get damped as a function of the probing time with a characteristic dephasing time $T_2$. Assuming that evolution of the field-sensor mixed state can be described in the Markovian  regime, which is the case in many AMO platforms, the effects of the noise on the time-evolution amounts to substituting $\ee^{\ii n\omega_0 (t-t_0)}\to\ee^{\ii n\omega_0 (t-t_0)}\ee^{-\frak{f}(\{x_j\})(t-t_0)/T_2}$ in the previous expressions, where  $\frak{f}(\{x_j\})$ is a particular function of the number and positions  of the probes. In some situations,  as occurs for the trapped-ion crystals~\cite{monz_14_qubit_entanglement} described below, these spurious couplings are mainly due to global fields, and $f(\{x_j\})=(\sum_j1)^2=n^2$, such that the visibility of the Ramsey parity signal decays  faster as the number of quantum sensors increases, limiting the advantage of this type of entangled quantum sensors in other contexts~\cite{entangled_q_Sensors_noise}. This sets a constraint into the proposed protocol, as only space-time coordinates fulfilling ${\rm max}\{|x_i^0-x_j^0|\}<\tau\ll T_{2}/n^2$  could be probed. 

To overcome this limitation, and given that the protocol already requires single-probe addressability, one may encode the sensors in a decoherence-free subspace by considering an entangled Neel-type initial state
\beq
\label{eq:partial_df_ghz}
\ket{\Psi_{\pm}(t_0)}=\ket{\Omega}\otimes\frac{1}{\sqrt{2}}\left(\prod_{{\bf x}\in X_{\rm e}}\sigma^1(t_0,{\bf x})\pm\prod_{{\bf x}\in X_{\rm o}}\sigma^1(t_0,{\bf x})\right)\ket{0_\sigma}.
\eeq
where $X_{\rm o}=\{{\bf x}_1, {\bf x}_{3}\cdots{\bf x}_{n-1}\}$, and $X_{\rm e}=\{{\bf x}_2, {\bf x}_{4}\cdots{\bf x}_{n}\}$.
 Additionally, the Schwinger sources~\eqref{eq:upgrade} must be modified to $J_\alpha(x)=J(x)\delta_{\alpha,3}/2$, and the Schwinger field~\eqref{eq:source_n_point} must become staggered  $J(x)=\sum_{i=1}^n\mathsf{J}_i(-1)^{i+1}\delta^D(x-x_i),$ which requires alternating field-sensor couplings. In this case, the parity signals  for each of the entangled Neel-type initial states $\mathsf{P}_{\pm} [J]= \bra{\Psi_{\pm}(t_0)}U_J^\dagger\prod_{{\bf x}\in X_{\rm s}}\sigma^1(t_0,	{\bf x})U_J\ket{\Psi_{\pm}(t_0)}$  lead to the following functional derivatives
\beq
\left.\frac{\delta^n\mathsf{P}{_\pm} [J(x)]}{\delta J(x_1)\cdots\delta J(x_n)}\right|_{J=0}=\half
G^{(n)}(x_1,\cdots,x_n)\pm{\rm c.c.},
\eeq
which directly yield the real ($+$) and imaginary ($-$) parts of the $n$-point propagator. The prescription to evaluate the functional derivatives would be similar as the one described above. In the ideal case, we have assumed that  $f(\{x_j\})=(\sum_j(-1)^j)^2=0$, such that no decoherence will affect the parity signals. In practice, as discussed in more detail below, there will be non-global components of the source-field coupling and other sources of noise  that will degrade the visibility of the parity oscillations, limiting the possible space-time points of the propagators that can be measured. We would also like to comment on a different strategy to combat the effect decoherence by combining the measurement scheme for the propagators~\eqref{two_point_propagator} with dynamical decoupling techniques (i.e. concatenated spin-echo sequences)~\cite{dyn_decoupling}. In the impulsive regime where the Schwinger sources are switched on/off very fast~\eqref{eq:source_n_point}, the spin echoes will only refocus the decohering  effect of the much slower fluctuating fields, but will not affect the signal that we aim to measure.

\subsection{Finite temperature and other interacting  QFTs}
\label{other_qfts}

So far, we have focused on a self-interacting bosonic QFT at $T=0$. As mentioned in the introduction, the connection to open problems in high-energy physics, such as the phase transition between the hadron gas and quark-gluon plasma in quantum chromodynamics, would require considering finite-$T$ regimes and other QFTs that include fermionic matter at finite densities coupled to  gauge fields. The question that we thus address in this subsection is whether the sensing scheme for the generating functional can be applied to finite temperatures, and generalized to other QFTs.

Let us start by discussing the finite-$T$ regime in the self-interacting Klein-Gordon QFT~\eqref{eq:int_lag}. The generating functional in this case becomes
\beq
\label{eq:generating_functional_finite_T}
\mathsf{Z}_T[J(x)]={\rm Tr}\left(\rho_{T}\mathsf{T}\left\{\ee^{\ii\int {\rm d}^DxJ(x)\phi_{H}(x)}\right\}\right),
\eeq
where $\rho_{ T}=\ee^{-\beta\int{\rm d}^dx\mathcal{H}}/{\rm Tr ( \ee^{-\beta\int{\rm d}^dx\mathcal{H}}})$ is the Gibbs state of the   QFT with Hamiltonian $\mathcal{H}$~\eqref{eq:Ham} at temperature $T=1/\beta$. By functional differentiation, and using Eq.~\eqref{eq:n-part_propagator}, one recovers the correct $n$-point Feynman propagators at finite temperature  $G^{(n)}={\rm Tr}\left(\rho_{T}\mathsf{T}\left\{\phi_{H}(x_1)\cdots \phi_{H}(x_n)\right\}\right)$.
 Such a functional can be inferred from the spin  parity oscillations of the quantum sensors, provided that the initial state is $\rho(t_0)=\rho_{T}\otimes\ket{\Psi(t_0)}\bra{\Psi(t_0)}$, where the initial state for the sensors corresponds to the GHZ state of  Eq.~\eqref{eq:ghz_all}. In this case, we are   assuming that  the self-interacting Gibbs state $\rho_T$ can be prepared dissipatively in the distant past, while the GHZ spin state is prepared in analogy to the $T=0$ case. Since the distant-future time-evolution operator is still given by Eq.~\eqref{eq:time_evolution}, one can directly prove that the finite-$T$ spin parity evolves as 
\beq 
\label{eq:full_parity_T}
\mathsf{P}_T[J]= {\rm Tr}\left(\!\rho(t_0) U_J^\dagger\prod_{{\bf x}}\sigma^1(t_0,{\bf x})U_J\!\right)=\half\ee^{\ii\int{\rm d}^Dx\delta\epsilon }\mathsf{Z}_T[J]+{\rm c.c.},
\eeq
and thus encodes the desired finite-$T$ generating functional.
From this expression, one can directly reproduce the previous results for the Feynman propagators, which now correspond to finite-$T$ time-ordered Green's functions. This can be generalized to initial states that are diagonal in the energy eigenbasis of the interacting    QFT, but not necessarily distributed according to the Boltzmann weights.

Let us now discuss the generalization of these ideas to other QFTs, such as  $N$-component scalar fields $\{\phi_a(x)\}_{a=1}^N$, which can be used to model the scalar Higgs sector in the Standard Model via an $O(N)$ Klein-Gordon QFT with $\lambda(\sum_a\phi_a^2(x))^2$ interactions.  Measuring the most generic generating functional of this QFT would require the same sensors but with couplings to each of the field components that can be switched on/off independently (i.e.\ different Schwinger functions $\{J^\alpha_a(x)\}_{a=1}^N$). However, for the symmetry-broken phase, it may suffice to use a single source coupled to one component which is singled out (Higgs component vs Goldstone
modes).   For the gauge-field sector of the Standard Model, the quantum sensors need to be coupled to  each gauge potential  $\{A_\mu^a(x)\}$, where $a\in\{1,\cdots,N_{\rm g}\}$ depends on the number of generators of the gauge group,  e.g. for the electromagnetic
field in $3+1$ dimensions it suffices  to consider four different source fields $\{J^\alpha_\mu(x)\}_{\mu=0}^3$ that can be switched on/off independently. The situation gets more complicated for the matter sector of the Standard Model, since these may require using also fermionic quantum sensors
instead, whose combined action together with standard sources could play
the role of the usual Grassmann Schwinger fields that appear in the
generating functional. We leave this possibility for a future work, and
instead comment on the possibility of using the protocol to measure
generating functionals where the Schwinger sources are coupled to fermion
bilinears, e.g.\ in the form of currents. This will be of relevance for
transport and linear response theory, in which transport properties can be
extracted from real-time correlators using Kubo relations. One example is
the electrical conductivity, which is of interest for a wide range of
systems, from graphene to the quark-gluon plasma.

\section{\bf Application to quantum simulators of QFTs}

In this section, we argue that AMO quantum simulators  are an ideal scenario  where to apply   our protocol  to measure the generating functional of a QFT. By exploiting quantum entanglement and  coherence, the quantum simulator can function as a  non-perturbative gadget that calculates the Feynman propagator, and thus the corresponding Feynman diagrams to  all orders in the interaction parameters. According to the introduction, we will need to put our findings in the generic context of  lattice field theories, which is addressed in Subsec.~\ref{lattice_sensors}. In Subsecs.~\ref{ions_phi4} and~\ref{ion_sensors}, we 
 discuss the direct connection of these  lattice-field theory concepts to  AMO quantum simulators based on crystals of trapped atomic ions. After outlining this connection, we describe in detail renormalization and the continuum limit of a generic scalar field theory in Subsec.~\ref{renormalization_lattice}, making connections to the  trapped-ion implementation that offer a practical view of this abstract topic.

\subsection{QFT and quantum sensors on the lattice}
\label{lattice_sensors}

In the following section, we will  focus on the AQS of interacting QFTs since, in principle, these simulators can be scaled  up  to the large sizes required to take  the continuum limit without the need of quantum error correction. From this perspective, we must consider lattice field theories in real time, where it is only the $d$-dimensional space which is discretized  on a lattice $ \Lambda_{\ell}=a\mathbb{Z}_N^d=\{{\bf x}:x_{\alpha}/a\in\mathbb{Z}_N,\forall\alpha=1,\cdots,d\}$ with $a$ being the lattice spacing, and $\mathbb{Z}_N=\{1,2,\cdots, N\}$~\cite{ham_lgt}. However, we note that the scheme could  be generalized to DQS, which could address the continuum limit by exploiting quantum-error correction to minimize the accumulated Trotter errors  and gate imperfections for increasing system sizes.

Once again, we will focus on the self-interacting scalar QFT, such that the field operator $\phi({\bf x})$ and its canonically-conjugate momentum  $\pi({\bf x})=\partial_t\phi({\bf x})$ are only defined for ${\bf x}\in\Lambda_\ell$, and fulfil $[\phi({\bf x}),\pi({\bf y})]=\ii\delta_{{\bf x},{\bf y}}/a^d$ which become the standard commutation relations $[\phi({\bf x}),\pi({\bf y})]=\ii\delta^d({{\bf x}-{\bf y}})$ in the continuum limit  $a\to 0$. To put the QFT~\eqref{eq:Ham} on a lattice~\cite{book_lattice}, we need to discretize the spatial derivatives of the Hamiltonian density, and substitute integrals by Riemann sums, such that the Hamiltonian of the lattice field theory reads
\beq
\label{eq:phi_4_lattice}
H=\!\!\sum_{{\bf x}\in\Lambda_\ell}\!\!a^d\!\left(\half\pi({\bf x})^2+\half(\nabla\phi({\bf x}))^2+\frac{m_0^2}{2}\phi({\bf x})^2+\mathcal{V}(\phi)\!\right)\!,
\eeq
where $(\nabla\phi({\bf x}))^2=\sum_{\alpha}\left(\phi({\bf x}+a{\boldsymbol{u}_\alpha})-\phi({\bf x})\right)^2/a^2$ is the sum of  forward differences along the   axes with unit vectors $\boldsymbol{u}_\alpha$.  The spatial lattice serves as a regulator for the QFT, as the high-energy modes  are cut-off by the finite lattice spacing, such that only momenta below the cut-off are allowed $|\boldsymbol{p}|\leq\Lambda_{\rm c}=2\pi/a$. As announced in the introduction,
 taking the continuum limit removes the   cut-off $\Lambda_{\rm c}\to\infty$, and one has to be careful with the UV divergences that appear in loop integrals  when $\mathcal{V}(\phi)\neq 0$~\cite{qft_book,book_lattice}. In this case, the bare mass  $m_0$ no longer coincides with the physical mass  $m$ of the particles, but  becomes instead a cutoff-dependent parameter $m_0(\Lambda_{\rm c})$  through a so-called renormalization process that shall be discussed in more detail below.

\begin{figure*}
\centering
\includegraphics[width=1.7\columnwidth]{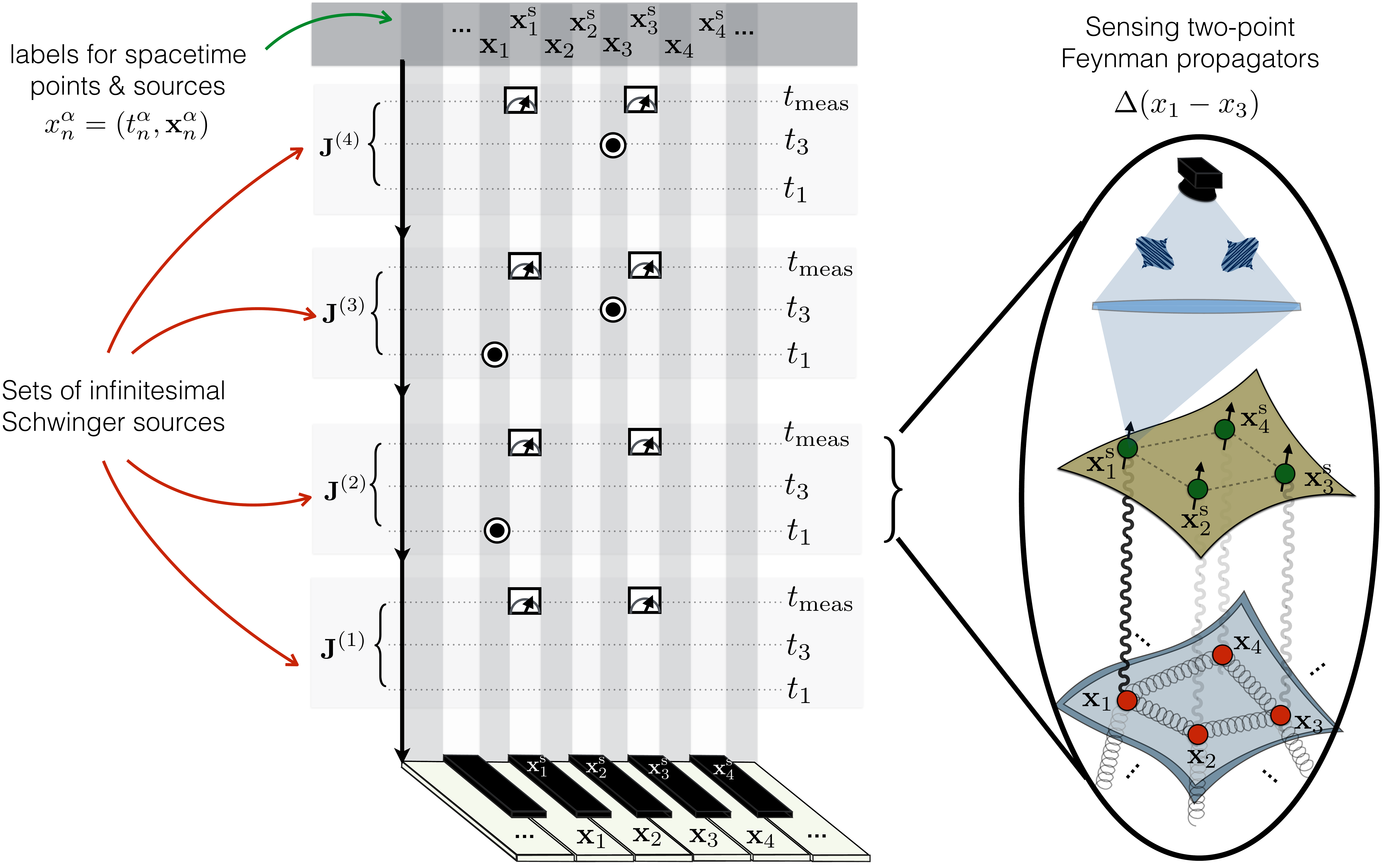}
\caption{ {\bf Schematic representation of the quantum sensing for Feynman propagators: } The different indexes for the lattice sites ${\bf x}\in\Lambda_{\rm s}$, as well the corresponding $\mathbb{Z}_2$ sensors  labelled by ${\bf x}^{\rm s}$, are mapped onto the keys of a piano. The set of pulse sequences  $\mathbf{J}^{(m)}$  that couple the sensors to the field~\eqref{eq:source_n_point} corresponds to a piano score that indicates the sequence of keys (sensors)  that must be pressed (coupled to the field) at different instants of time to produce a melody (spin-parity measurement) that encodes the relevant information about the Feynman propagators. }
\label{fig_field_sensing_piano}
\end{figure*}

Let us now introduce the lattice $\mathbb{Z}_2$ Schwinger sources~\eqref{eq:upgrade} by attaching a spin-1/2 quantum sensor $\sigma^\alpha_{{\bf x}}$ to each lattice point ${\bf x}\in\Lambda_\ell$, and defining a lattice Schwinger field $J^\alpha_{{\bf x}}(t)$. Accordingly, we have to supplement the above  Hamiltonian of the lattice field theory with 
\beq
\label{eq:z2_schwinger_sources}
H\to H_J=H+H_\sigma-\sum_\alpha\sum_{{\bf x}\in\Lambda_\ell} a^dJ^\alpha_{{\bf x}}(t)\phi({\bf x})\sigma^\alpha_{{\bf x}},
\eeq 
where the dynamics of the sensors is governed by 
\beq
\label{eq:sensor_ham}
 H_\sigma=\sum_{{\bf x}\in\Lambda_\ell} a^d \delta\epsilon(\sigma^0_{{\bf x}}-P_{{\bf x}}),
 \eeq
  and  $P_{{\bf x}}$ projects onto the ground state of the sensor at lattice site ${\bf x}\in\Lambda_\ell$.
Considering a  Ramsey-type scheme $J^\alpha_{{\bf x}}=J_{{\bf x}}(t)(\delta_{\alpha,0}-\delta_{\alpha,3})/2$, the time-evolution operator~\eqref{eq:time_evolution} on the lattice can be expressed as 
\beq
\label{eq:time_evolution_lattice}
U_J(t,t_0)=\mathsf{T}\left\{\ee^{-\ii(t-t_0){H}_0}\right\}\mathsf{T}\left\{\ee^{+\ii \int_{t_0}^t {\rm d}t'\sum_{{\bf x}\in\Lambda_\ell}a^d J_{{\bf x}}(t)\phi_{\rm H}(t',{\bf x})P_{{\bf x}}}\right\},
\eeq 
where $H_0=H+H_{\sigma}$ describes the uncoupled evolution of the  self-interacting lattice field and the $\mathbb{Z}_2$ sensors.
Considering an initial  maximally-entangled state for the lattice   sensors 
\beq
\ket{\Psi(t_0)}=\ket{\Omega}\otimes\frac{1}{\sqrt{2}}\left(\prod_{{\bf x}\in\Lambda_\ell}\sigma^0_{{\bf x}}+\prod_{{\bf x}\in\Lambda_\ell}\sigma^1_{{\bf x}}\right)\ket{0_\sigma},
\eeq
we find that  the corresponding spin-parity observable $\mathsf{P}[J,a]= \bra{\Psi(t_0)}U_J^\dagger\prod_{{\bf x}\in\Lambda_\ell}\sigma^1_{{\bf x}}U_J\ket{\Psi(t_0)}$ can be expressed as  
\beq
\mathsf{P}[J,a]=\half\ee^{\ii(t-t_0) \sum_{\boldsymbol{x}\in\Lambda_\ell}a^d \delta\epsilon }\mathsf{Z}[J,a]+{\rm c.c.},
\eeq
where we have introduced the lattice generating functional
\beq
\label{eq:lattice_gf}
\mathsf{Z}[J,a]=\bra{\Omega}\mathsf{T}\left\{\ee^{\ii\int_{t_0}^t {\rm d}t'\sum_{{\bf x}\in\Lambda_\ell}a^dJ_{{\bf x}}(t')\phi_{\rm H}(t',{\bf x})}\right\}\ket{\Omega}.
\eeq
The corresponding Feynman propagators $G^{(n)}_{{\bf x}_1,\cdots {\bf x}_n}(t_1,\cdots,t_n)$ can be obtained by functional differentiation, as described in the continuum version~\eqref{eq:functional_derivatives_parity}, 
where  one must consider again   $t-t_0>{\rm max}\{|t_i-t_j|,\hspace{1ex}\forall\hspace{0.5ex} i,j=1,\cdots ,n\}$ to approximate the remote-past to distant-future conditions. We recall that a set ${\mathbf{J}}^{(m)}$ of point-like sources~\eqref{eq:source_n_point} would be required, such that  one can  reconstruct the discretization of the functional derivatives by  the set of measured parities.

This lattice version offers a very vivid image of our quantum sensing apparatus as a piano (see Fig.~\ref{fig_field_sensing_piano}). Let us label the  $|\Lambda_\ell|$ lattice sites with an integer that maps each site to a particular key of a piano. The list $\mathbf{J}^{(m)}$, which describes the sequence of pulses that couple the sensors to the field~\eqref{eq:source_n_point}, can be interpreted as a piano score that indicates the sequence of keys (sensors)  that must be pressed (coupled to the field) at different instants of time to produce a melody (spin parity) that encodes the relevant information about the Feynman propagators.

Following~\cite{book_lattice}, the lattice generating functional~\eqref{eq:lattice_gf} in the non-interacting limit, $\mathcal{V}(\phi)=0$, becomes  $\mathsf{Z}_0[J,a]={\rm exp}\{-\half\int{\rm d}x^0\int{\rm d}y^0\sum_{\boldsymbol{x},\boldsymbol{y}\in\Lambda_\ell}a^{2d}J( x)\Delta_0({ x}-{ y}, a)J( y)\}$, where 
\beq
\Delta_0({ x}-{y},a)=\!\!\int \frac{{\rm d}p^0}{2\pi}\sum_{{\bf p}\in {\rm BZ}}\frac{\ii\ee^{\ii{ p( x- y)}}}{(p^0)^2-m_0^2-\sum_\alpha \left(\frac{2}{a}\sin\left(\frac{a}{2}p_\alpha\right)\right)^2},
\eeq
is the  single-particle Feynman propagator, and  we have introduced the Brillouin zone ${\rm BZ}=[-\frac{\pi}{a},\frac{\pi}{a})^{\times^d}$. From this expression, the corresponding propagator in momentum space 
\beq
\Delta_0( p,a)=\frac{\ii}{(p^0)^2-m_0^2-\sum_\alpha \big(\frac{2}{a}\sin\left(\frac{a}{2}p_\alpha\right)\big)^2},
\eeq
 has a well-defined continuum limit. Removing the lattice cut-off,  this propagator  coincides with that  of the free scalar  Klein-Gordon QFT $\lim_{a\to 0}\Delta_0(p,a)= \ii/(p^2-m_0^2)$~\cite{qft_book}, where  $p^2 = (p^0)^2- {\bf p}^2$.  Note that the   pole of the propagator at ${\bf p}^2=0$, which determines  the physical mass $m$ of the scalar particle, coincides in this case with  the bare mass $m_0$ of the original field theory~\eqref{eq:int_lag}.   

As noted below Eq.~\eqref{eq:phi_4_lattice}, the situation is  more involved when $\mathcal{V}(\phi)\neq0$, since  the  bare parameters of the theory must depend on the  cut-off to cure the  UV divergences. The particular cut-off-dependence of the bare parameters is determined by requiring that the  physical observables at the length scale of interest are not modified  when the number of  high-energy modes, describing fluctuations  at much smaller length scales, is increased in the continuum limit   $\Lambda_{\rm c}^{-1}\to 0$. Since $a$ (or $\Lambda_{c\rm }^{-1}$) is a length (or inverse energy) scale,
and hence not dimensionless, taking the continuum limit should always be
understood in the sense that $\xi/a\to \infty$. Here, $\xi$ sets the
 relevant length scale of interest in such a way that physical quantities
become independent of the underlying lattice structure. 

We will discuss this point in more detail  below, but let us first introduce a particular AMO platform that can be used as an AQS of a self-interacting scalar QFT on the lattice. Regarding the lattice counterpart of the sensing protocols for other continuum QFTs discussed in Subsec.~\ref{other_qfts}, a similar approach to the one presented in this section would hold for N-component scalar fields and fermion fields with bilinear sources. On the other hand, extending our sensing protocol to lattice gauge fields is an open question that deserves further studies, especially in view of the recent progress towards the quantum simulation of lattice gauge theories~\cite{dqs_lattice_gauge_theories,dqs_qft_gauge_U(1)_qlinks,dqs_lattice_gauge_theories,dqs_qft_SU(2)_qlinks,aqs_qft_gauge_U(1)_qlinks,aqs_QED,aqs_qft_gauge_U(1)_qlinks,aqs_qft_fermions_U(1)_qlinks,aqs_qft_fermions_SU(2)_qlinks,aqs_qft_fermions_U(1)_qlinks,aqs_qft_fermions_SU(2)_qlinks,zohar_review}.

\subsection{Trapped-ion  quantum simulators of the $\lambda\phi^4$ QFT}
\label{ions_phi4}

The possibility of trapping  atomic ions by electromagnetic fields has allowed to test the predictions of quantum mechanics at the single-atom level~\cite{ions_rmp,wineland_nobel}. After the seminal work by Cirac and Zoller~\cite{cz_gates}, it was understood that operating with several ions would allow for quantum information processing~\cite{ions_qc_review}, turning trapped ions into a very promising route towards quantum error correction~\cite{color_code_ions}. Prior to the development of a large-scale fault-tolerant quantum computer based on  trapped ions, one may exploit the experimental setup for quantum simulations~\cite{qs_ions}. As argued in the introduction, with few notable exceptions of DQS~\cite{lgt_ions}, the experimental emphasis has been placed on the quantum simulation of condensed-matter problems~\cite{QS_trapped_ions}.
However, as discussed in this section, trapped ions also have the potential of becoming useful AQS of relativistic QFTs in a high-energy physics context.

The motion of a system of $N$ trapped atomic ions of mass $m_{\rm a}$ and charge $e$ can be described  by the Hamiltonian
 \begin{equation}
 \label{eq:trapped_ions_motion}
 H_m=\sum_{i=1}^N\sum_{\alpha=x,y,z}\left(\frac{1}{2m_{\rm a}}p_{i\alpha}^2+\frac{1}{2} m_{\rm a}\omega_{\alpha}^2r_{i\alpha}^2\right)+\frac{e_0^2}{2}\sum_i\sum_{j\neq i}\frac{1}{|{\bf r}_i-{\bf r}_j|},
 \end{equation}
where we have introduced the position $r_{i\alpha}$ and momentum $p_{i\alpha}$ operators fulfilling  $[r_{i\alpha},p_{j\beta}]=\ii \delta_{i,j}\delta_{\alpha,\beta}$, and the effective trapping frequencies  $\{\omega_{\alpha}\}_{\alpha=x,y,z}$ in the pseudo-potential approximation~\cite{ions_rmp}. Here,  $e_0^2=e^2/4\pi\epsilon_0$ is expressed in terms of  the vacuum permittivity $\epsilon_0$, and we have set $\hbar=1$, which is customary in AMO physics since energies are then given by the frequency of the electromagnetic radiation used to excite a particular transition observed in spectroscopic measurements. 

 As a result of the competition between the Coulomb repulsion and the trap confinement, the ions can self-assemble in  Coulomb crystals of different geometries when the temperatures get sufficiently low~\cite{zigzag_ions}. In this article, we shall be interested in linear and zigzag crystal configurations, which are routinely obtained in linear Paul traps~\cite{zigzag_spectroscopy} and, more recently, also    in  a combination of a  Paul trap and an optical lattice~\cite{ion_crystal_optical_lattices}, which shall be referred to as a sub-wavelength Paul trap. In addition, the recent experiments showing the crystallization of ion rings in segmented ring traps~\cite{ion_ring} could also explore different crystal configurations. 
 
 In the harmonic-crystal approximation, one considers small vibrations around the equilibrium positions ${\bf r}_i={\bf r}_i^0+\sum_\alpha q_ {i,\alpha}{\bf e}_\alpha$, and obtains a model of coupled harmonic oscillators that leads to  the vibrational normal modes of the Coulomb crystal~\cite{modes_ion_crystal}. This approximation, however, cannot account for the motional dynamics of the ions close to a structural transition between different crystalline structures. In particular, when $\omega_y\gg\omega_x,\omega_z$, a structural change between a linear ion chain and  a zigzag ladder occurs as one lowers  $\omega_x$ below a critical value $\omega_{\rm c}$ via a  second-order phase transition~\cite{spt_ions_molecular_dynamics}.  This phase transition can be understood by an effective Landau model~\cite{landau_zigzag}, which identifies the transverse zigzag distortion where neighbouring ions vibrate  in anti-phase as a soft mode. For $\omega_x<\omega_{\rm c}$, the transverse phonons condense in  a different ladder  structure by spontaneously breaking a $\mathbb{Z}_2$ inversion symmetry. Not only is this theory in accordance with previous static predictions~\cite{spt_ions_molecular_dynamics}, but also serves as the starting point  for studies of   non-equilibrium dynamics of the  crystal across the phase transition~\cite{kibble_zurek_ions}.

\begin{figure*}
\centering
\includegraphics[width=1.7\columnwidth]{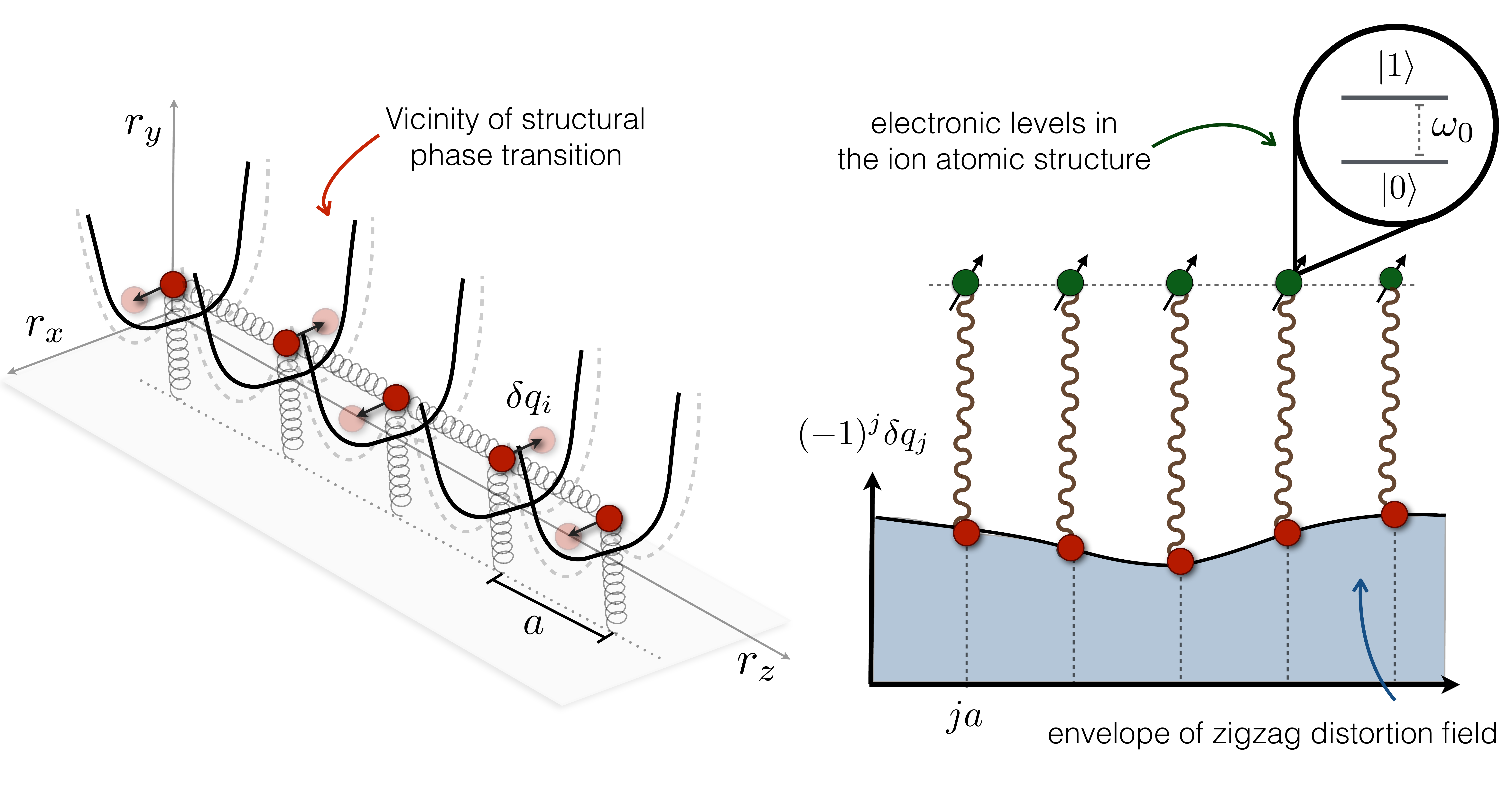}
\caption{ {\bf Effective $\lambda \phi^4$ QFT for ion crystals and quantum sensing scheme: } (left) In the vicinity of the linear to zigzag structural phase transition of a trapped-ion crystal, the transverse zigzag vibrations yield the soft mode that contains the universal properties of the transition. (right) The slowly-varying envelop of the zigzag distortion~\eqref{eq:zigzag_distortion} allows to develop a gradient expansion that leads to the  effective  $\lambda\phi^4$ QFT. By exploiting two electronic levels of  the ions, and a state-dependent dipole force~\eqref{eq:state-dependent_force}, one can infer the generating functional  of the QFT. }
\label{fig_field_sensing_ions}
\end{figure*}

An effective low-energy theory for the linear-to-zigzag  transition can be  derived as follows, both for ion rings~\cite{effective_phi_4} and inhomogeneous linear crystals~\cite{spin_peierls_ions}. Let us rewrite the equilibrium positions as ${\bf r}_i^0=a\tilde{\bf r}_i^0$, where $a$ is a relevant length scale in the problem. For  the sub-wavelength Paul traps or for ring traps, $a$ is the uniform lattice spacing, whereas for linear Paul traps where the crystals are inhomogeneous, $a=(e_0^2/m_{\rm a}\omega_z^2)^{1/3}$ is simply a  length scale with the order of magnitude of the average lattice spacing. We know from the previous discussion that the low-energy physics will be governed by excitations around the soft zigzag mode, which corresponds to momentum $k_{\rm s}=\pi/a$ in a ring trap (see Fig.~\ref{fig_field_sensing_ions}). In analogy with other  problems in condensed matter, see e.g.~\cite{affleck_notes}, one puts a cut-off around this momentum, considering only low-energy excitations that should capture the long-distance physics.
This amounts to 
rewriting the zigzag distortion as $q_ {j,x}=\ee^{\ii k_{\rm s}  ja}\delta q_ {j}$,  where   $\delta q_{j}$ is  a displacement that is slowly-varying on the scale of the lattice spacing which only contains the modes near $k_{\rm s}$. This can be generalized to situations without the periodicity of the ring  by simply defining 
\beq
\label{eq:zigzag_distortion}
q_ {j,x}=(-1)^j\delta q_ {j}.
\eeq 
A gradient expansion $\delta q_{j}\approx \delta q_{i}+(\tilde{\bf r}_j^0-\tilde{\bf r}_i^0)\partial_{i+1}\delta q_{i}$, where $\partial_{j}\delta q_{i}=(\delta q_{i}-\delta q_{j})$ fulfils $|\partial_{j}\delta q_{i}|\ll\delta q_{i}$ due to its slowly-varying condition,  yields the following Hamiltonian
 \begin{equation}
 \label{eq:ham_ion_crystal}
 H_{m}\approx\!\sum_i\!\left(\!\frac{m_{\rm a}}{2}\left(\partial_t \delta q_{i}\right)^2\!+\!\frac{\tilde{k}_{i}}{2}(\partial_{i+1}\delta q_{i})^2+\frac{k_i}{2}\delta q_{i}^2\!+\frac{u_i}{4}\delta q_{i}^4\!\right)\!.
 \end{equation}
Here, we have introduced a local spring constant and self-interaction coupling for each transverse displacement
\begin{equation}
\label{eq:mic_1}
k_{i}=m_{\rm a}\omega_x^2\left(1-\half\kappa\zeta_i(3)\right),\hspace{1ex} u_i=\frac{3}{4a^2}m_{\rm a}\omega_x^2\kappa\zeta_i(5),
\end{equation} 
where  $\zeta_i(n)=\sum_{l\neq i}[(-1)^i-(-1)^l]^{n-1}|\tilde{{\bf r}}_i^0-\tilde{\bf r}_l^0|^{-n}$, and $\kappa=e_0^2/m_{\rm a}\omega_x^2a^3$ is a dimensionless constant. Additionally, the spring constants  between  neighbouring displacements are 
\begin{equation}
\label{k_app}
\tilde{k}_{i}=m_{\rm a}\omega_x^2\sum_{l\neq i}\frac{(-1)^{l+i+1}\kappa}{2|\tilde{\bf r}_i^0-\tilde{\bf r}_l^0|}.
\end{equation}

The Hamiltonian in Eq.~\eqref{eq:ham_ion_crystal}  already resembles the lattice field theory of a $D=1+1$  Klein-Gordon QFT~\eqref{eq:phi_4_lattice}, where the underlying ion crystal plays the role of the  
$d=1$  lattice  
\beq
\label{eq:ion_lattice} 
\Lambda_\ell=\{{\rm x}: {\rm x}/a=\tilde{\bf r}_i^0, \forall i=1,\cdots N\}.
\eeq
We thus specialize to $D=1+1$ dimensions, in which, as noted below Eq.~\eqref{eq:int_lag}, the engineering dimension of the scalar field is $d_\phi=(D-2)/2=0$. In order to define the correct scalar field operators, one has to pay special attention  to the different system of units in equations~\eqref{eq:phi_4_lattice} and~\eqref{eq:ham_ion_crystal}.  Essentially, we need to identify  the  speed of sound that will play the role of the effective speed of light in the relativistic QFT. Since the scalar field must be dimensionless, we start by defining the  following lattice operators 
\beq
\label{eq:field_displacement}
\tilde{\phi}({\rm x})=\frac{1}{a}\delta q_i,\hspace{2ex}\tilde{\pi}({\rm x})=m_{\rm a}\partial_t\delta q_i,
\eeq
which show the desired commutation relations $[\tilde{\phi}({\rm x}),\tilde{\pi}({\rm y})]=\ii \delta_{\rm x,y}/a$. The lattice Hamiltonian~\eqref{eq:ham_ion_crystal} can then be expressed as $H_{m }=H_0+V$, where we have introduced 
\beq
H_0=\sum_{{\rm x}\in\Lambda_\ell}\!\!a\hspace{0.2ex} \!\left(\frac{\tilde{\pi}({\rm x})^2}{2m_{\rm a}a}+\frac{\tilde{k}_{ i}a^3}{2}(\nabla\tilde{\phi}({\rm x}))^2\right),
\eeq	
and used the operator $\nabla\tilde{\phi}({\rm x})=\left(\tilde{\phi}({\rm x}+a{{u}_x})-\tilde{\phi}({\rm x})\right)/a$. This part  can be rewritten in terms of 
\beq
H_0=\sum_{{\rm x}\in\Lambda_\ell}\!\!a\hspace{0.2ex}\frac{c_{\rm x}}{2} \!\left(\frac{\tilde{\pi}({\rm x})^2}{K_{\rm L,x}}+K_{\rm L,x}(\nabla\tilde{\phi}({\rm x}))^2\right),
\eeq
where we have introduced an effective sound velocity 
  \beq
 	\label{eq:par1}
 c_{\rm x}=a\sqrt{\frac{\tilde{k}_i}{m_{\rm a}}},\hspace{2ex}
 \eeq
 which has the correct dimensions $[c_{\rm x}]=[\tilde{k}_ia^2]^{1/2}\cdot [m_{\rm a}]^{-1/2}=(\mathsf{ML^{2}T^{-2}}\cdot\mathsf{M^{-1}})^{1/2}={\mathsf{LT}}^{-1}$. Additionally, we have also introduced the so-called stiffness or Luttinger parameter
 \beq
 \label{eq:lutt_parameter}
  {K}_{\rm L, x}=a^2\sqrt{\tilde{k}_i m_{\rm a}},
 \eeq
 which appears in the theory of bosonization and controls the power-law decay of correlations in Luttinger liquids~\cite{cm_book}. Reintroducing Planck's constant, $  {K}_{\rm L, x}=\frac{a^2}{\hbar}\sqrt{\tilde{k}_i m_{\rm a}}$, this Luttinger parameter turns out to be dimensionless ${K}_{\rm L, x}=([\tilde{k}_ia^2]/[\hbar ^2/m_{\rm a}a^2])^{1/2}=(\mathsf{ML^{2}T^{-2}}/\mathsf{ML^{2}T^{-2}})^{1/2}=1$.
 In order to arrive at the standard definition of a $\lambda\phi^4$ QFT on the lattice~\eqref{eq:phi_4_lattice}, we perform an additional rescaling of the lattice field operators that preserves the commutation relations
 \beq
 \label{eq:fields_ions}
 {\phi}({\rm x})=\sqrt{{K}_{\rm L, x}}\tilde{\phi}({\rm x}),\hspace{2ex}  {\pi}({\rm x})=\frac{1}{\sqrt{{K}_{\rm L, x}}}\tilde{\pi}({\rm x}).
 \eeq
This leads to the desired lattice field theory
 \beq
H_0=\sum_{{\rm x}\in\Lambda_\ell}\!\!a\hspace{0.2ex}\frac{c_{\rm x}}{2} \!\left({\pi}({\rm x})^2+(\nabla{\phi}({\rm x}))^2\right),
\eeq
which yields the desired QFT of a  1+1 free massless scalar boson $H_0=\int{\rm dx}\frac{c_{\rm x}}{2} \!\left({\pi}({\rm x})^2+(\partial_{\rm x}{\phi}({\rm x}))^2\right)$  in the continuum limit  $a\to 0$. Note that, as a consequence of the inhomogeneous lattice spacing in a linear Paul trap, all these parameters have inhomogeneities around the edges of the ion chain,  while they become  constants for ring traps and sub-wavelength Paul traps,  where the lattice spacing is homogeneous.

In addition to these terms, the remaining part of the lattice Hamiltonian~\eqref{eq:ham_ion_crystal} yields a mass term and a self-interaction of the scalar field 
 \beq
 \label{eq:V_latt}
 V=\sum_{{\rm x}\in\Lambda_\ell}\!\!a\left(\frac{m_{0,\rm x}^2}{2}\phi({\rm x})^2+\frac{{\lambda}_{\rm x}}{4!}{{\phi}}({\rm x})^4\right).
 \eeq
 Here, we have introduced the bare mass and coupling strength 
 \beq
  	\label{eq:par3}
  {m}^2_{0,\rm x}=\frac{k_ia}{K_{\rm L,x}},\hspace{2ex}{\lambda}_{\rm x}=\frac{6u_ia ^3}{K_{\rm L,x}^2},
 \eeq 
 which fulfil $[ a{m}^2_{0,\rm x}]= [ a\lambda_{\rm x}]=\mathsf{ML^2T}^{-2}$ after taking into account the  lattice spacing $a$ from the lattice sum in Eq.~\eqref{eq:V_latt}, and thus display   the expected units of energy.
 In  Landau's mean-field theory, ${m}_{0,\rm x}^2<0$, ${\lambda}_{\rm x}>0$ signals a phase transition where $\langle {\phi}({\rm x})\rangle\neq 0$ is achieved by spontaneously breaking  the $\mathbb{Z}_2$  symmetry ${\phi}({\rm x})\to-{\phi}({\rm x})$ in the effective lattice field theory~\eqref{eq:phi4_ions}. This is exactly in agreement with previous estimates  of the linear-to-zigzag phase transition in ion crystals, both in  homogeneous and inhomogeneous cases. However, the mean-field  approach predicts a wrong scaling behavior in the vicinity of the critical point, which could be tested experimentally with the protocol  presented in this work.

 In the context of relativistic QFTs~\eqref{eq:int_lag}, we should recover Lorentz invariance in the continuum limit.  This can be achieved for the whole ion crystal in ring traps or sub-wavelength Paul traps,  or by restricting to  the homogeneous bulk of the crystal in a linear Paul trap.  In these cases, we can   set the corresponding natural units $c=1$, such that the low-energy Hamiltonian governing the linear-to-zigzag instability in ion crystals becomes equivalent to the $D=1+1$ lattice  Klein-Gordon field~\eqref{eq:phi_4_lattice} with quartic interactions 
 \beq
  \label{eq:phi4_ions}
 H_m=\!\sum_{{\rm x}\in\Lambda_\ell}\!a\left(\half\pi({\rm x})^2+\half(\nabla\phi({\rm x}))^2+\frac{m_0^2}{2}\phi({\rm x})^2+\frac{\lambda}{4!}\phi({\rm x})^4\right).
\eeq
Note that with all these definitions, we have made sure that the classical mass dimensions $d_\phi=0$, while the couplings have $d_{m_0^2}=2=d_{\lambda}$. Finally, we also note that in numerical lattice simulations and  formal renormalization group (RG) calculations, one typically defines dimensionless couplings $\tilde{m}_0^2=m_0^2a^2$ and $\tilde{\lambda}=\lambda a^2$. In the so-called lattice units, the lattice constant disappears from the above Hamiltonian, such that taking the continuum limit corresponds to modifying the dimensionless couplings. As remarked at the end of the previous  subsection, taking the continuum limit does not require changing the actual inter-ion distance, but  is instead achieved by setting  these dimensionless couplings close to a critical point where $\xi/a\to \infty$, and  physical quantities
become independent of the underlying lattice structure.

According to this discussion,  trapped-ion crystals can be used as  AQSs of the lattice $\lambda\phi^4$ QFT~\eqref{eq:phi4_ions}, where the fields~\eqref{eq:fields_ions} are proportional  to the zigzag displacement~\eqref{eq:zigzag_distortion} via Eq.~\eqref{eq:field_displacement}. The proportionality parameter, as well as the bare mass and self-interaction strength of the QFT are expressed in terms of microscopic parameters~\eqref{eq:mic_1}-\eqref{k_app} via Eqs.~\eqref{eq:par1}-\eqref{eq:lutt_parameter} and \eqref{eq:par3}. We note that this approach differs from a  non-canonical transformation introduced in~\cite{hbar_zigzag}, which yields a similar Hamiltonian~\eqref{eq:phi4_ions} after a particular rescaling of  Eq.~\eqref{eq:ham_ion_crystal}. However,  an effective  Planck  constant $\hbar$ depending on the model parameters must be introduced  to maintain the required commutation relations. This leads to important differences in the renormalization with respect to the standard approach for the $\lambda\phi^4$ theory, which shall be discussed  below.

As advanced in the introduction, the usefulness of an AQS does not only depend on the accuracy with which it behaves according to the model of interest, e.g. a self-interacting scalar QFT, but also on the measurement strategies to extract the  relevant properties of this simulated model. In our trapped-ion scenario, the position of the ions $\langle {\bf r}_i\rangle$ is routinely measured by driving a transition between two electronic levels and collecting  the spontaneously-emitted photons in a camera~\cite{ions_rmp}, such that the expectation value $\langle \phi({\rm x})\rangle$ could be inferred from the above relations. This can be used to locate the critical point of the $\mathbb{Z}_2$ phase transition when a vacuum expectation value $\langle \phi({\rm x})\rangle \neq 0$ is developed, which would require and accurate  measurement of the zigzag ion positions~\cite{zigzag_spectroscopy}. However, 
in the symmetry-broken phase, the zigzag crystal will also experience micromotion (i.e.\ additional fast oscillations synchronous with the driving fields of the Paul trap) that go beyond the pseudo-potential approximation, such that other spectroscopic observables can be modified with respect to the static situation~\cite{zigzag_spectroscopy}. In the present  context,  the pseudo-potential approximation is used to derive Eq.~\eqref{eq:phi4_ions}, and it would thus be safer for the accuracy of the AQS to perform experiments in the symmetry unbroken phase, where these standard fluorescence measurements cannot be used to determine the properties of the QFT.  For instance, if one is interested in simulating a massive scalar particle with the AQS, one would like to know how the bare mass gets renormalized as a consequence of the self-interactions, or how a collection of  massive scalar particles would scatter off each other  due to these interactions. In the following section, we show that the protocol to measure the generating functional $\mathsf{Z}[J,a]$~\eqref{eq:lattice_gf}, which contains the information about all of these properties, and  has an experimental realization that is feasible with state-of-the art control over trapped-ion crystals.

\subsection{Trapped-ion  sensors for the generating functional}
\label{ion_sensors}

The Hamiltonian~\eqref{eq:trapped_ions_motion} describes the motional degrees of freedom of a collection of trapped ions. Additionally, the ions have an internal atomic structure with its own independent dynamics. We can exploit such internal degrees of freedom as the quantum sensors introduced in Sec.~\ref{sec:sensors_qft} (see Fig.~\ref{fig_field_sensing_ions}). 

We will consider external laser beams that only couple to a pair of such internal states   $\{\ket{0_i},\ket{1_i}\}$, which  have a transition of frequency $\omega_0$. The Hamiltonian governing this internal dynamics is simply
$
H_{in}=\sum_{i=1}^N\omega_0(\sigma^0_i-P_i)$,
where $\sigma^0_i=\mathbb{I}=\ket{0_i}\bra{0_i}+\ket{1_i}\bra{1_i}$, and $P_i=\ket{0_i}\bra{0_i}$ is the projector onto the lowest-energy internal state. This Hamiltonian is directly equivalent to the  quantum-sensor Hamiltonian~\eqref{eq:sensor_ham} using the crystal as the underlying lattice~\eqref{eq:ion_lattice}, such that 
\beq
\label{eq:h_sensors_ions}
H_{in}=\sum_{{\rm x}\in\Lambda_\ell}\! a\hspace{0.1ex} \delta\epsilon(\sigma^0_{{\rm x}}-P_{{\rm x}}), \hspace{2ex} \delta\epsilon=\frac{\omega_0}{a}.
\eeq

 In order for these electronic levels to act as quantum sensors of the QFT generating functional, we need to induce a coupling of the form~\eqref{eq:z2_schwinger_sources}, such that these probes act as the $\mathbb{Z}_2$ Schwinger sources introduced in Sec.~\ref{sec:sensors_qft}. 
We consider the so-called state-dependent dipole forces~\cite{sd_force}, which can be obtained from a pair of laser beams of frequency $\omega_{\rm L,1},\omega_{\rm L,2}$ that couple the internal state $\ket{0}_i$ off-resonantly to an auxiliary excited state from the atomic level structure. Using selection rules~\cite{cats_ions},  and working in the far off-resonant regime, the laser-ion coupling can be expressed as a crossed-beam ac-Stark shift 
\beq
\label{eq:li_int}
H_{l-i}=\sum_i \frac{\Omega_{\rm L}}{2}P_i\ee^{\ii (\Delta {\bf k}_{\rm L}\cdot{\bf r}_i-\Delta\omega_{\rm L} t)}+{\rm H.c.},
\eeq
where we have introduced a two-photon Rabi frequency $\Omega_{\rm L}$, and the wave-vector (frequency) difference of the laser beams
 $\Delta{\bf k}_{\rm L}=\Delta{\bf k}_{\rm L,1}-\Delta{\bf k}_{\rm L,2}$ ($\Delta{\omega}_{\rm L}=\Delta{\omega}_{\rm L,1}-\Delta{\omega}_{\rm L,2}$). After expressing the ion position operator in terms of the vibrations ${\bf r}_i={\bf r}_i^0+\sum_\alpha q_ {i,\alpha}{\bf e}_\alpha$, one sees directly that if the overlapping beams propagate along $\Delta {\bf k}_{\rm L}||{\bf e}_x$, the radiation will couple to the desired zigzag distortion~\eqref{eq:zigzag_distortion}. Moreover, a Taylor expansion in the Lamb-Dicke regime, $|\langle\Delta {\bf k}_{\rm L}\cdot{\bf e}_x q_{i,x}\rangle|\ll 1$, shows that the leading-order contribution from the laser-ion interaction~\eqref{eq:li_int}, for $|\Omega_{\rm L}|\ll \Delta\omega_{\rm L}\sim \omega_x$, will be a state-dependent dipole force that excites the zigzag distortion when the internal state of the ions is in $\ket{0_i}$, namely
 \beq
 \label{eq:state-dependent_force}
 H_{l-i}=\sum_i g_i(t)P_i\delta q_i,\hspace{1ex} g_i(t)=\Omega_{\rm L}(\Delta {\bf k}_{\rm L}\cdot{\bf e}_x a)(-1)^i\sin{\Delta\omega_{\rm L} t}.
 \eeq
We will also assume that the laser beams can be  split into individual addressing beams that couple to  any ion in the crystal, such that $\Omega_{\rm L}\to\Omega_{{\rm L},i}(t)$ can be controlled individually, e.g. switched on/off,   by controlling the  intensity of each of the addressing beams as achieved experimentally in~\cite{ind_addressing_raman}. Using this expression in combination with Eqs.~\eqref{eq:phi4_ions} and~\eqref{eq:h_sensors_ions}, we arrive at the desired lattice field theory with $\mathbb{Z}_2$ Schwinger sources~\eqref{eq:z2_schwinger_sources}, namely
\beq
H_J=H_{m}+H_{\rm in}-\sum_\alpha\sum_{{\bf x}\in\Lambda_\ell} aJ^\alpha_{{\rm x}}(t)\phi({\rm x})\sigma^\alpha_{{\rm x}},
\eeq
where we have introduced the source fields
\beq
J^\alpha_{{\rm x}}(t)=\frac{ J_{{\rm x}}(t)}{2}(\delta_{\alpha,0}-\delta_{\alpha,3}),\hspace{2ex} J_{{\rm x}}(t)=\frac{g_i(t)}{\sqrt{K_{\rm L,x}}}.
\eeq
Note that, by using the dimensional analysis of the previous section, the source fields also have the desired mass dimensions $d_J=2$.

Provided that one can prepare the initial  state of the ions according to Eq.~\eqref{eq:ghz_all}, and measure the parity observable in Eq.~\eqref{eq:full_parity},  it  becomes possible to implement the protocol  presented in Sec.~\ref{sec:sensors_qft}, inferring the generating functional $\mathsf{Z}[J,a]$ of the particular trapped-ion $\lambda\phi^4$ QFT.  For the $T=0$ case, the state preparation would rely on an adiabatic evolution that starts far away from the structural phase transition, and utilizes laser cooling to prepare a  state very close to the vacuum of the transverse vibrations. Then, the trap  parameters would be adiabatically modified by approaching the critical point of the linear-to-zigzag transition, but remaining in the symmetry-preserved phase. For the $T\neq0$ case, one would perform laser cooling directly in the interacting regime, during a time that is large enough so that the motional degrees of freedom thermalize. Then, the internal state has to be prepared in a GHZ state, which can be accomplished using gates mediated by the  phonons that are not involved in the structural phase transition~\cite{entangled_ions}. We remark that the high fidelities already achieved in the experimental preparation of large GHZ states~\cite{monz_14_qubit_entanglement} make  trapped ions a very promising AMO setup for the implementation of this proposed protocol.

Before closing this subsection, let us note that the simplified protocols of Sec.~\ref{sec:sensors_qft} to measure any Feynman propagator could also be implemented in this trapped-ion scenario, provided that one has the aforementioned addressability in the laser-ion couplings~\cite{ind_addressing_raman}. In such case, the  state in Eq.~\eqref{eq:partial_ghz} or~\eqref{eq:partial_df_ghz}  could be prepared along similar lines, and the required switching of instantaneous sources to estimate the functional derivatives~\eqref{eq:functional_derivatives_parity} would also be available.  The measurement corresponds to a multi-spin correlation function of the type that is routinely measured through the state-dependent fluorescence of the ions~\cite{ions_rmp}. Prior to  driving the cycling transition and collecting the emitted photons, one should  apply a global single-qubit rotation by driving the so-called carrier transition~\cite{ions_rmp}. We note that the measurements have to be repeated for different values of the field-sensor couplings  to infer the propagators via the discretized derivatives~\eqref{two_point_propagator}. During these  additional repetitions, one must avoid slow drifts in the microscopic trapped-ion parameters. An advantage in this regard is that our proposal focuses  on the propagator of the vibrations, which will be 1-2 orders of magnitude faster than  experiments on the propagation of spin excitations in effective spin-spin models with trapped ions~\cite{propagation_spin_excitations_ions}, where analogous measurements are typically done.

\subsection{Renormalization and the continuum limit}
\label{renormalization_lattice}

As advanced in the sections above, using the lattice generating functional $\mathsf{Z}[J,a]$~\eqref{eq:lattice_gf} to learn about the continuum QFT~\eqref{eq:int_lag} requires letting $a\to 0$, and  removing the lattice cut-off $\Lambda_{\rm c}\propto a^{-1}\to\infty$. This continuum limit must be performed without affecting the physical observables at the length scale of interest. Note also that the  Schwinger sources should be spaced at the same
physical distance as the 'continuum limit' is taken. For instance, in the context of the trapped-ion quantum simulator~\eqref{eq:phi4_ions}, such an observable will be  the parity operator~\eqref{eq:parity_generating}, which encodes the information about the Feynman propagators~\eqref{eq:n-part_propagator} and thus  the physical mass $m$ of the scalar particles. In this case, the relevant length scale for the scalar fields~\eqref{eq:fields_ions} is set by the envelope of the zigzag distortion~\eqref{eq:zigzag_distortion}, which varies on a much larger scale than the lattice spacing. In the generic situation, we can safely send $a\to 0$ without altering the long-wavelength phenomena, but we must ensure that  our calculations  will not suffer from possible UV divergences as  further high-energy modes are included by this process. In practice, this requires allowing the bare couplings of the theory $\{g_i\}$,  e.g. $\{m_0^2,\lambda\}$ in Eq.~\eqref{eq:phi4_ions}, to flow with the lattice cut-off $\{g_i(\Lambda_{\rm c})\}$  in such a way that one stays on the 'line of constant physics'. In the AQS, this would mean that the value of the microscopic parameters, which control the effective lattice parameters such as the bare mass,  have to be tuned to particular values in order to obtain the renormalized physical mass of the particles at the scale of interest, which will be independent of the cut-off and  different from the bare mass. The renormalization group  is essential  to understand this flow and, with it, the nature of such a  continuum limit~\cite{rg_hollowod}. 

At the UV limit $\{g_i(\infty)\}$, the resulting  QFT must belong to the so-called critical surface, i.e.\ the couplings must lie at the domain of attraction of a fixed point of a transformation   that changes the cut-off scale.  To preserve the physics at the length-scale of interest, one has to fix a one-parameter set of field theories with different cut-offs $\{g_i(\Lambda_{\rm c})\}$ that connects to such a well-defined UV limit.  This is achieved by specifying the relevant couplings $\{g_i^{\rm r}(\Lambda_{\rm c})\}\in\{g_i(\Lambda_{\rm c})\}$ that take us away from the critical surface as one moves  from the UV  towards the infra-red (IR) $\Lambda_{\rm c}\to 0$, approaching  thus the length-scale of interest. The difficulty lies in identifying the possible RG fixed points and relevant couplings of a  particular field theory. In this regard, the scalar QFT~\eqref{eq:int_lag} with self-interactions  $\mathcal{V}(\phi)=\sum_ng_{2n}\phi^{2n}/(2n)!$ and $D=4$ yields a very instructive scenario where the RG machinery can be developed in perturbation theory~\cite{rg_hollowod,wilson_rg}. Typically, one starts from the so-called Gaussian fixed point, where $g_{2n}(\infty)=0$, and shows that it suffices to consider the flow of $g_{2}(\Lambda_{\rm c})$ and $g_4(\Lambda_{\rm c})$ to understand the continuum limit. This follows from simple dimensional analysis,  since the so-called anomalous dimensions of the fields vanish at this fixed point, allowing one to realize that the higher-order couplings $\{g_{2n}(\Lambda_{\rm c})\}_{n>2}$ are all irrelevant, i.e.\ decrease as one moves towards the IR. At one loop in perturbation theory, $g_{2}(\Lambda_{\rm c})$ remains a relevant coupling, while $g_{4}(\Lambda_{\rm c})$  becomes irrelevant. Therefore, unless a different RG fixed point exists, the lattice regularization of the scalar QFT in   $D=4$ only has a trivial non-interacting continuum limit. Using the so-called  $\epsilon$ expansion,  which allows for non-integer dimensions $D=4-\epsilon$, it is possible to find a non-trivial fixed point that would allow for an interacting and massive QFT in the continuum, the so-called Wilson-Fisher fixed point at finite $g_{2}(\infty)\neq 0,g_{4}(\infty)\neq 0$. However, this fixed point  exists only for  $\epsilon>0$ and thus  $D<4$, suggesting the triviality of the lattice scalar QFT in   $D=4$~\cite{wilson_rg,triviality_lattice}. 

To go beyond the perturbative RG, numerical lattice simulations based on Monte Carlo methods  become very useful~\cite{book_lattice}. The general strategy of lattice field theory simulations is to set the bare lattice couplings in the vicinity of a quantum critical point, where  the correlation length $\xi\to\infty$ diverges, and one expects to recover the universal features of the QFT in the continuum limit $a\to 0$. In our context, $g_{2}(\Lambda_{\rm c})$ and $g_4(\Lambda_{\rm c})$ must be set in the vicinity of the $\mathbb{Z}_2$ quantum phase transition, which should be controlled by the scale-invariant fixed point of the RG transformation. The renormalized mass can be extracted from  the numerical computation of propagators, whereas the renormalized interactions can be obtained from 
 susceptibilities.  This approach corroborates the triviality of the lattice scalar QFT in $D=4$ in non-perturbative regimes~\cite{triviality_monte_carlo}.  

In the $D=2$ limit, which is the case of interest for the trapped-ion quantum simulator~\eqref{eq:phi4_ions}, the need of non-perturbative schemes is even more compelling. In this case, applying the above perturbative RG calculation around the Gaussian fixed point would show that all of the couplings $\{g_{2n}(\Lambda_{\rm c})\}_{n\geq 1}$ are relevant~\cite{rg_hollowod}, thus questioning the validity of the truncation implicit in Eq.~\eqref{eq:ham_ion_crystal} that is used to derive the effective QFT~\eqref{eq:phi4_ions} from the microscopic Hamiltonian~\eqref{eq:trapped_ions_motion}. In fact, in 1+1 dimensions, the field operators for a free scalar QFT have non-vanishing anomalous dimensions even in the absence of interactions, such that the simple dimensional analysis around the Gaussian fixed point is no longer valid.  In this case, the tools of conformal field theory would be required to understand the RG flow of perturbations around   the scale-invariant fixed point of a free scalar boson, as occurs for the sine-Gordon model~\cite{cm_book}. However,  the particular perturbations of our self-interacting scalar QFT do not have simple conformal/scaling dimensions, and thus do not allow for a simple  analytical approach. Accordingly, the existence of non-perturbative numerical methods  becomes even more relevant in this situation. Recent results for this 1+1 scalar  $\lambda\phi^4$ QFT, either based on Monte Carlo~\cite{phi_4_1d_montecarlo} and real-space renormalization group~\cite{phi_4_1d_dmrg_mps} methods on the lattice, or  Hamiltonian truncation methods in a finite volume~\cite{phi_4_1d_ham_truncation}, have shown that  the continuum limit of this QFT is controlled by a non-trivial fixed point corresponding to the Ising conformal field theory. These works show  the power of the lattice approach to solve non-perturbative questions of the continuum QFT, such as the precise location of the $\mathbb{Z}_2$ quantum phase transition, i.e.\ the critical value of $ \lambda/m^2$ where the scalar field acquires a vacuum expectation value. At a fundamental level, they also imply that perturbations $\{g_{2n}(\Lambda_{\rm c})\}_{n>2}$ around this fixed point, which are generated in the implicit RG process of  looking into long-wavelength phenomena, are irrelevant. This   justifies thus the validity of our truncation leading to Eq.~\eqref{eq:phi4_ions}. The hope of this manuscript is to show that, exploiting the proposed protocol to infer the full generating functional $\mathsf{Z}[J,a]$ of a QFT, trapped-ion AQS working in the vicinity of the linear-to-zigzag structural  transition will serve as an alternative non-perturbative tool to explore such   QFT questions.

\section{\bf Conclusions and outlook}

 In this work, we have presented a  protocol to infer the normalized generating functional of  a QFT by measuring a particular interferometric observable  through a collection  two-level quantum sensors. Generalizing the notion of Schwinger fields  to serve simultaneously as  sources and  probes of the excitations of a quantum field, we have  exploited the entanglement of the  quantum sensors to show that a collective Ramsey-type response of the    sensors contains all the information about the QFT generating functional. This,  in turn, encodes  in a compressed manner  the relevant information of the interacting QFT (i.e.\ approximating functional derivatives by combining several measured responses can be used  to decompress any $n$-point Feynman propagator, and thus any possible scattering or non-equilibrium real-time process). We have argued that this protocol finds a very natural realization in AQS on the lattice, and we have considered a trapped-ion realization of the $\lambda\phi^4$ QFT as a realistic example where experimental techniques can be applied to implement the generalized Schwinger sources, and infer the generating functional from resonance-fluorescence images. In this case, by performing experiments in the vicinity of the linear-to-zigzag structural phase transition, the trapped-ion AQS can in principle address non-perturbative questions regarding the nature of the fixed point that controls the  QFT obtained in the continuum limit. During the completion of our work, S. P. Jordan et {\it al.}~\cite{dqs_gen_func} have shown that the algorithm for DQS of scattering in scalar QFTs~\cite{dqs_qft_phi4} can be modified to obtain also the generating functional. Moreover, they argue that a particular instance of the generating functional, for a certain functional dependence of the source fields, cannot be efficiently estimated with any classical algorithm. It would be very interesting to study if similar complexity arguments can be carried over onto our AQS, which we believe presents an opportunity to measure the generating functional using current trapped-ion technology.

In general, understanding real-time dynamics of quantum fields, either in or out of equilibrium,
is required  in a wide range of physical applications. One
important question in relativistic theories is far-from equilibrium
dynamics and thermalization~\cite{NONEQ}. This is relevant e.g. for the end of
inflation and preheating in the early universe~\cite{PRE}, or for the evolution
during the early stages of heavy-ion collisions, resulting in the
quark-gluon plasma (QGP) created at the Large Hadron Collider at CERN, or
the Relativistic Heavy Ion Collider at BNL~\cite{QGP-rev}{. In the former, the
efficiency of particle production and transport of energy across different length
scales determines the reheating temperature, whereas in the latter case
the very creation of a thermal QGP depends on the ability of the highly
non-equilibrium initial gluon fields to thermalize~\cite{THERM}. Since this
dynamics  takes place manifestly in real-time, it is often treated within
classical approximations that are only valid for highly occupied modes~\cite{HIC}. It would
hence be  of the utmost interest to study similar, yet simplified, dynamical questions relevant
for these situations using the protocol outlined in our work, and analyze e.g.
the role of non-thermal fixed points~\cite{FP} using quantum dynamics in real
time.

In thermal equilibrium, the information about  spectral functions and other
real-time correlators  is also of interest. Even though they are related~\cite{LeBellac} to standard
Euclidean correlation functions computable in lattice field theory, the
analytical continuation from Euclidean to real time (or from Matsubara to
real frequency) is a non-trivial process. The interest here lies e.g. in
quasi-particle properties and other spectral features, such as thermal
masses and widths, or in more ambitious questions related to hydrodynamic
structure and transport at long wavelengths~\cite{TRANS}. It would be interesting to explore if similar questions  can be
addressed extending the present protocol for measurements of thermal current-current
correlators in real time.

{\it Acknowledgements.--}
We thank S. P. Kumar for useful discussions.  This work has been supported by STFC grant ST/L000369/1. A.B.  acknowledges support from Spanish MINECO Projects FIS2015-70856-P,  and CAM regional research consortium QUITEMAD+.   G. A. is supported by the Royal Society and the Wolfson Foundation. This work has been supported by STFC Grant No. ST/L000369/1.


\begin{references}
\bibitem{comment_free_fermions}
Provided that one takes care of the so-called fermion doublers~\cite{nielsen_nonomiya} that appear in the naive discretization of a fermionic Dirac field.
\end{references}
\end{document}